\documentclass[conference]{IEEEtran}

\usepackage{times,epsfig,graphicx,wrapfig,bbm,color}
\usepackage{bm, amssymb, amsmath, amsfonts, bbm}

\usepackage{cite}
\usepackage{url}
\def\BibTeX{{\rm B\kern-.05em{\sc i\kern-.025em b}\kern-.08em
    T\kern-.1667em\lower.7ex\hbox{E}\kern-.125emX}}
    
\usepackage{color}
\definecolor{red}{rgb}{1,0.2,0.2}
\definecolor{green}{rgb}{0.2,1,0.5}
\definecolor{blue}{rgb}{0,0,1}
\definecolor{lightblue}{rgb}{0.3,0.5,1}

\newcommand {\N} {{\rm I\kern-2.5pt N}}
\newcommand {\R} {{\rm I\kern-2.5pt R}}

\newcommand{\beqa}{\begin{eqnarray}}
\newcommand{\eeqa}{\end{eqnarray}}
\newcommand{\beqan}{\begin{eqnarray*}}
	\newcommand{\eeqan}{\end{eqnarray*}}
\newcommand{\beq}{\begin{equation}}
\newcommand{\eeq}{\end{equation}}

\newcommand{\bfl}{\begin{flushleft}}
	\newcommand{\efl}{\end{flushleft}}
\newcommand{\bgnv}{\begin{verbatim}}
	\newcommand{\ednv}{\end{verbatim}}

\newcommand{\myeq}{& \hspace{-0.1in} = & \hspace{-0.1in}}

\newcommand{\myarr}{\begin{array}{lll}}
	\newcommand{\indicate}[1]{{\bf{1}} \left\{#1\right\}}

	\newcommand{\bx}{{\bf x}}

\newcommand{\bepsilon}{\boldsymbol{\epsilon}}

\newcommand{\bitem}{\begin{itemize}}
	\newcommand{\eitem}{\end{itemize}}
\newcommand{\benum}{\begin{enumerate}}
	\newcommand{\eenum}{\end{enumerate}}

\begin{document}

\title{Detection of Performance Interference Among Network Slices in 5G/6G Systems}

\author{
\IEEEauthorblockN{Van-Sy Mai}
\IEEEauthorblockA{\textit{NIST}\\
vansy.mai@nist.gov}
\and
\IEEEauthorblockN{Richard J. La}
\IEEEauthorblockA{\textit{Univ. of Maryland}\\
hyongla@umd.edu}
\and
\IEEEauthorblockN{Tao Zhang}
\IEEEauthorblockA{\textit{NIST (Ret.)}\\
taozhang1@yahoo.com}
\and
\IEEEauthorblockN{Bin Hu}
\IEEEauthorblockA{\textit{NIST}\\
bin.hu@nist.gov}}

\maketitle

\begin{abstract}
Recent studies showed that network slices (NSs), which are logical networks supported by shared physical networks, can experience service interference due to sharing of physical and virtual resources. Thus, from the perspective of providing end-to-end (E2E) service quality assurance in 5G/6G systems, it is crucial to discover possible service interference among existing NSs in a timely manner and isolate the potential issues before they can lead to violations of service quality agreements. We study the problem of (a) detecting service interference among NSs in 5G/6G systems and (b) identifying misbehaving NSs and other affected NSs, only using E2E key performance indicator measurements, and propose new algorithms. Our numerical studies demonstrate that, even when the service interference among NSs is weak to moderate, provided that a reasonable number of measurements are available, the proposed algorithms can correctly identify most of shared resources that can lead to service interference among the NSs that utilize the shared resources and misbehaving NSs that can cause potentially adverse service interference and affected NSs. 
\end{abstract}

\section{Introduction}	\label{sec:Intro}


Network slicing is a new technology that allows a single physical network to be shared by multiple logical networks, called {\em network slices} (NSs)~\cite{Foukas2017, Rost2017}. This allows different NSs to be set up to support heterogeneous traffic classes over the same physical network infrastructure in a dynamic manner to meet different needs for diverse users. 


In 5G/6G systems, the NSs are envisioned to be supported by virtual network functions (VNFs), rather than dedicated hardware, to improve network flexibility. A VNF may share physical or virtual resources with other VNFs and consequently network traffic flows of different NSs may share physical or virtual resources. As a result, what happens in one NS (e.g., changes of traffic volume and traffic routes; and security compromises) may adversely affect other NSs. We refer to this as {\em NS interference} or {\em service interference}. 

Since NSs may traverse end to end (E2E) across multiple autonomous systems (ASs)---access and core networks---interference among them may occur anywhere throughout a network where they share any physical or virtual resources at any protocol layer. Furthermore, the set of resources shared by any two NSs at any given time also depends on dynamically changing factors, such as the network routes of the traffic flows in these NSs at the time. 

The ability to detect E2E service interference and assess how the interference may impact each NS will be essential to delivering the promise of 5G and future networks to support E2E differentiated services, especially E2E service quality assurance, to support diverse applications.

Existing studies on interference in a network have focused primarily on local interferences – interferences inside one part of a network (e.g., inside radio access networks) or caused by specific network components or technology. Examples include interferences between radio channels \cite{zambianco2020interference, xu2013rethinking}, between virtual machines (VMs) \cite{amri2017inter} or virtual network functions (VNFs) \cite{zhang2019adaptive}, and inside a cloud computing system \cite{jain2017systematic, kambadur2012measuring}. These local interference measurement methods alone, however, fail to provide adequate pictures of E2E service interference that are necessary for understanding and controlling interference to assure E2E service quality.

This naturally leads to the following important question: {\em Is it possible to detect or even predict potential interference among NSs, using only E2E key performance indicator (KPI) measurements, such as E2E delays and packet drops?} Having the capability to detect potential service interference among existing NSs can help the service providers isolate the cause of potential problems in a timely manner and avoid violations of E2E quality assurance, and improve the quality of user experience. To the best of our knowledge, our study is the first to investigate the problem of discovering possible service interference among NSs using only E2E KPI measurements.

A key challenge arises from the fact that the internal operations of each AS (such as an access network) are unlikely to be known to or controllable by other ASs (such as core networks). Such internal AS operations include how physical and virtual resources---such as VFNs, computing capacities, network routes, and communication bandwidth---are assigned at any given time, often dynamically, at different protocol layers of each NS across an AS. As a result, no single entity in the network may possess complete knowledge of how (physical or virtual) resources are shared among NSs at any time end-to-end across multiple ASs. For this reason, to answer the above question, we need to first identify (a) the shared resources, and which NSs are sharing each resource, and (b) any misbehaving NSs that cause potentially damaging service interference to other NSs and those affected by the misbehavior. This is the focus of the current study. As we illustrate in the following sections, there exist several technical challenges to addressing this important problem.
\\ \vspace{-0.12in}

\underline{\bf Summary of main contributions:} 
We first present preliminary experimental results obtained from our small custom testbed that is used not only to confirm the presence of service interference between NSs, but also to demonstrate the extent of service interference when virtual and physical resources are shared even in a small network supporting two NSs (Section~\ref{sec:Testbed}). This motivates our investigation in this paper. 

We propose a novel algorithm for identifying, using only the E2E measurements, shared resources in a network and how NSs sharing the resources may interfere with each other (Section~\ref{sec:Proposed-shared-resources}). The algorithm is based on factor analysis (FA)~\cite{BarthKnott} and consists of three phases: in the first phase, it builds an interference graph using pairwise correlations in E2E measurements among NSs, which are measured using the Spearman's rank correlation coefficients~\cite{Daniel}. The second phase generates a list of maximal cliques in the interference graph, which are then used in the third and final phase to identify all cliques that represent shared resources with their sharing NSs.

Second, we put forth an algorithm for finding misbehaving NSs that can cause service interference via shared resources (Section~\ref{sec:proposed-misbehaving-NSs}). To this end, it first computes the expected KPI measurements for each NS using the output of the first algorithm and then uses the difference between the expected values and measurements to cluster the NSs. Once misbehaving NSs are discovered, we can find the NSs that are affected by them using the outcome from the first algorithm. 

We carry out extensive numerical studies focusing on scenarios where the service interference among NSs is weak to at most moderate, with the Pearson correlation coefficients among resource-sharing NSs ranging mostly from 0.05 to 0.25 (Section~\ref{sec:Numerical}). Our numerical studies illustrate that, even with weak to moderate service interference among the NSs, the proposed algorithm can correctly detect most of the shared resources in a network and sharing NSs, provided that sufficient measurements are available. Furthermore, the second algorithm can reliably find most of misbehaving NSs  (with the missed detection rate below 10 percent) with very small false alarm rates. We also demonstrate that the proposed algorithms are robust to measurement noise; even when the standard deviation of measurement noise is 50 percent of correct value, our algorithms experience only minor degradation in performance. Finally, we provide numerical results to show that the Spearman's rank correlation coefficient is better suited for detecting correlations between interfering NSs than the Pearson correlation coefficient. 

\subsection{Related work}
    \label{subsec:Related}
    
Our problem is related to network tomography (or network inference) that makes use of measurements at the network edge to examine the internal network characteristics and performance~\cite{caceres1999, coates2000, duffield2001, malekzadeh2013, Tsang2003, ziot2001} or to discover network topology~\cite{bestavros2002, coates2002a, duffield2001a, duffield2002a}. The first part of our problem also requires identifying the set of shared resources and is thus related to network topology identification. The second part of our problem estimates the congestion level within each NS. Therefore, it has some resemblance to the internal network inference problem. 

At the same time, there are major differences between our problem and both network topology identification and internal network inference. For example, most of the existing literature on network topology identification focuses on tree-like topologies by considering traffic from a single source. Also, studies on internal network characteristics and performance primarily deal with estimating link-level traffic inside the network or its parameters. Furthermore, to the best of our knowledge, FA has not been applied to network tomography, but is a very natural approach to our problem and numerical studies demonstrate that it is very effective. 

FA employed in our algorithms can be viewed as a dimensionality reduction technique and has some similarity to other linear dimensionality reduction techniques. For example, principal component analysis (PCA), a well-known dimensionality reduction technique, is related to FA; both attempt to approximate measurements using a small number of latent variables that serve as weights for associated vectors. A key difference between FA and PCA is that while PCA aims to find a small number of {\em orthogonal} vectors that capture most of variability in measurements, FA does not look for orthogonal vectors. As we will explain, the vectors we aim to identify (called factor loadings) are often required to be non-orthogonal, suggesting that PCA is not appropriate for our problem. We will elaborate on this in Section~\ref{sec:Proposed-shared-resources}. 

Initial findings of our work has been reported in \cite{CISS2025}; it contains a description of the algorithm for finding shared resources in the network and provides some numerical results that demonstrate its effectiveness. However, it does not include the second algorithm designed to identify misbehaving NSs or its evaluation (Section~\ref{sec:proposed-misbehaving-NSs}). Moreover, it only considers the measurements without symmetrization (Section~\ref{subsec:SymVsAsym}). Finally, it does not examine the robustness of the first algorithm, which is studied in Section~\ref{sec:Numerical}. 
\\ \vspace{-0.1in}

The remainder of this paper is structured as follows: Section~\ref{sec:Preliminaries} introduces a short summary of FA and delineates symmetrization of measurements. Section~\ref{sec:Setup} outlines the problem formulation. Section~\ref{sec:Proposed-shared-resources} describes our proposed solution approach and algorithm for identifying shared resources in the network, followed by an exposition on our algorithm for finding misbehaving NSs in Section~\ref{sec:proposed-misbehaving-NSs}. Section~\ref{sec:Numerical} presents the numerical studies. Finally, Section~\ref{sec:Conclusion} includes the conclusions and future directions.

\section{Preliminaries}
    \label{sec:Preliminaries}

In this section, we first provide a short description of FA, which is used in our algorithm discussed in Section~\ref{sec:Proposed-shared-resources}. Then, we explain how we pre-process the KPI measurements before providing them as input to our algorithms.

\subsection{Factor Analysis}
    \label{subsec:FA}

FA is a well known statistical technique for describing the variability in observed measurements with the help of latent variables. Since typically the number of latent variables is much smaller than the dimension of measurements, it is considered a dimensionality reduction technique. Here we provide a brief overview of FA. In Section~\ref{sec:Proposed-shared-resources}, we describe how it is used to identify shared resources and the NSs sharing respective resources.
    
Suppose that ${\bf X}$ is an $n \times p$ matrix containing observations, where each column contains the values of one observed variable. Let $\overline{\bx} = [\overline{x}_1 \ \cdots \ \overline{x}_p]$ be the row vector containing the mean of the $p$ observed variables. The premise is that, although $p$ may be large, a smaller number, say $q$, of latent variables is responsible for generating each row of ${\bf X}$. Specifically, there is a set of $q$ (row) vectors $\{{\bf L}_1, \ldots, {\bf L}_q\}$, called {\em factor loadings}, such that each row of ${\bf X}$ minus $\overline{\bx}$ can be written as a linear combination of these $q$ vectors plus some noise: 
\beqa
{\bf X}_k - \overline{\bx} = \sum_{r=1}^q f_{k,r} {\bf L}_r + \bepsilon_k, \quad k = 1, 2, \ldots, n \ . 
	\label{eq:FA1}
\eeqa 
The weights $f_{k,1}, \ldots, f_{k,r}$ are the latent variables called {\em common factors}, and $\bepsilon_k$ represents the noise called {\em specific factors}. In a matrix form, we have
\beqan
{\bf X} = {\bf F L} + {\bf E} + \overline{{\bf X}} \ , 
\eeqan
where ${\bf F}$ is an $n \times q$ matrix whose $k$-th row is ${\bf F}_k = (f_{k,r} : r = 1, \ldots, q)$ containing the common factors for ${\bf X}_k$, ${\bf L}$ is a $q \times p$ matrix whose rows are the factor loadings  ${\bf L}_r$, $r = 1, \ldots, q$, ${\bf E}$ is an $n \times p$ matrix whose $k$-th row is $\bepsilon_k$ comprising the specific factors of ${\bf X}_k$, and $\overline{\bf X}$ is an $n \times p$ matrix whose rows are all equal to the mean vector $\overline{\bx}$.  We refer a reader interested in a more detailed discussion on FA to a monograph~\cite{BarthKnott}. Our proposed algorithm described in Section~\ref{sec:Proposed-shared-resources} utilizes FA to identify a set of shared resources and sharing NSs.

\subsection{Data Symmetrization for Factor Analysis}  
    \label{subsec:SymVsAsym}

In this subsection, we discuss how we pre-process the measurements before providing them as input to our proposed algorithm based on FA explained above. Our numerical studies reported in Section~\ref{sec:Numerical} demonstrate that pre-processing the measurements yields a significant improvement in performance when the number of measurements is small or interference between NSs is weak.

$\bullet$ {\bf Motivation:} Recall from \eqref{eq:FA1} that, in order to apply the FA to compute the factor loadings, common factors, and specific factors, we first need to subtract the mean vector $\overline{\bf x}$ from the observations ${\bf X}_k$, $k = 1, \ldots, n$, and compute the centered observations. One issue with subtracting the mean vector $\overline{\bf x}$ is that, when the number of observations $n$ is small, the estimated mean vector has large variance and may not be accurate. As a result, subtracting a noisy/inaccurate mean vector from the observations can distort the statistical properties of the centered observations and, consequently, they may behave differently than the correct centered observations. For this reason, when FA is applied to the statistically incorrect centered measurements in our problem, it can fail to correctly identify shared resources and misbehaving NSs, degrading the performance of the algorithm in our problem.  

$\bullet$ {\bf Proposed solution:} As explained above, a main source of performance degradation is the large variance in the estimated mean vector for small $n$. A possible way to reduce the variance of the mean vector is to augment the observations so that the mean vector becomes more predictable and the difference provides more reliable centered observations to which we can apply the FA. 

To this end, we first compute for each observed variable $i$ its minimum value and subtract the minimum value from the values of the variable in the $i$-th column of ${\bf X}$: let $x^i_{\min} := 
\min\{X_{k,i} : k = 1, \ldots, n\}$ and $\check{X}_{k,i} = X_{k,i} - x^i_{\min}$, $i = 1, \ldots p$, and $k = 1, \ldots, n$. Then, we construct an augmented matrix ${\bf X}^s$ by appending $-\check{\bf X}$ at the end of $\check{\bf X}$. In other words, ${\bf X}^s$ is a $2n \times p$ matrix whose top $n$ rows contain $\check{\bf X}$ and the bottom $n$ rows have $-\check{\bf X}$. 

It is clear that the mean of ${\bf X}^s$ is equal to the zero vector by construction. Therefore, when we apply FA to ${\bf X}^s$, we do not need to subtract a noisy mean vector. Moreover, from the viewpoint of FA, the same observations are presented twice -- one with the plus sign and the other with the minus sign. Consequently, a good approximation of observations in the top $n$ rows obtained by FA will also lead to an equally good approximation of augmented observations in the bottom $n$ rows. We will call ${\bf X}^s$ the {\em symmetrized} observations. 

\section{Problem Statement}	\label{sec:Setup}

We are interested in designing algorithms for detecting potential service interference between NSs based only on E2E KPI measurements, such as E2E delays, and for identifying misbehaving NSs that could cause adverse performance impact on other NSs. The interference can be caused at any shared resources---physical and virtual---including communication links, VMs, CPUs, GPUs, cache, and memory. 

In principle, a network should be configured to ensure sufficient isolation among the NSs so that an NS does not suffer noticeable performance degradation as a result of network traffic dynamics (e.g., traffic congestion) in other NSs. However, recent studies showed that VMs sharing physical resources can experience non-negligible interference. As a result, when multiple NSs share one or more resources, congestion at a shared resource may adversely affect the performance of other NSs that share the congested resource.

For example, when two or more NSs share a VNF (e.g., a user plane function (UPF)) or when VMs supporting multiple VNFs share CPUs or memory, the processing delays experienced by packets belonging to an NS can be affected by the traffic load of other NSs. Similarly, even when VNFs are allocated to different VMs and CPUs on a shared physical machine, they may still experience interference via cache or memory access even though the characteristics and extent of interference may vary depending on the types of shared resources.

\begin{figure}[h]
\centerline{
	\includegraphics[width=3.2in]{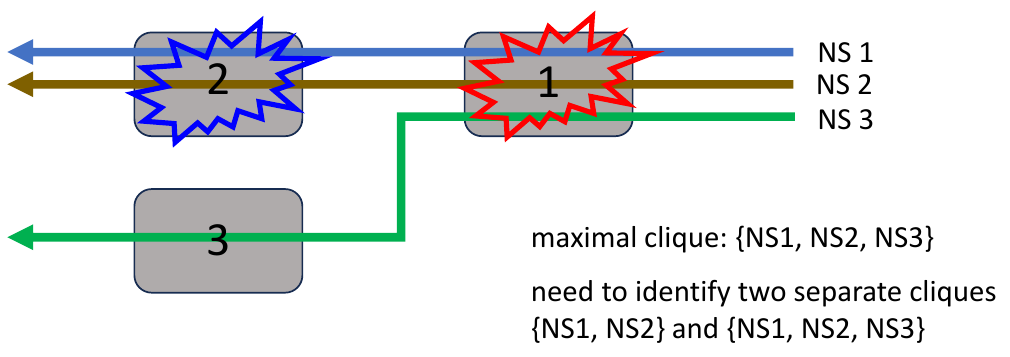}
}
\caption{Example with two shared resources.}
\label{fig:example}
\end{figure}

Correctly identifying different sources of potentially problematic service interference and interfering NSs due to resource sharing is challenging for several reasons. For instance, consider the example shown in Fig.~\ref{fig:example}. Here two resources are shared by three NSs: resource 1 is shared by all three NSs, and resource 2 is shared only by NSs 1 and 2. Since all three NSs share resource 1, the measurements will likely show correlations among all three NSs. Thus, it may be difficult to determine whether or not these correlations are caused by a single shared resource or by more than one resources that are shared by different subsets of NSs on the basis of E2E measurements. As explained in the following section, we address this challenge with the help of FA. It captures the variations in the E2E measurements, which are introduced by different shared resources, with the help of latent variables that summarize the state of shared resources (e.g., congestion levels).

\section{Service Interference in Testbed}
    \label{sec:Testbed}

In this section, we provide experimental results that demonstrate service interference among NSs when they share either physical or virtual resources. In particular, our experiments suggest that when NSs share virtual resources, service interference can be pronounced, leading to potentially significant challenges in providing E2E QoS assurance. 

$\bullet$ {\bf Testbed:} 
The testbed consists of three rack-mounted servers. One server (type 1) has an Intel server board S2600BP that includes two Intel Xeon Silver 4110 processors with 8 cores $@$ 2.10 Ghz, 48 GB DDR4 RAM, an Intel C620 PCH Integrated 10 Gigabit Ethernet controller, and an Intel Omni-Path NFI 100 series adapter.\footnote{Any mention of a commercial product is for information only and does not imply an endorsement or recommendation by NIST.} This machine is used to support data plane (DP) VNFs. The other two machines (type 2) have a 6039P-TXRT server board with 2 Xeon Gold 6240 processors with 18 cores $@$ 2.6 GHz, 256 GB DDR4 RAM, and two Intel 10 Gigabit X550T NICs. They are used to run the application servers and user equipments (UEs) and to support control plane VNFs. 

Our core network implementation is based on the Open5GCore platform \cite{open5gcore}. Constant traffic flows are generated using iperf2 \cite{iperf}, and time-varying traffic patterns are generated using D-ITG (Distributed Internet Traffic Generator) that can support stochastic processes at the packet level \cite{ditg}.

$\bullet$ {\bf Setup:}
In our experiments, there are a total of 12 flows handled by two NSs. Flows 1 through 8 are handled by NS 1, and flows 9 through 12 are handled by NS 2. Flows 1 through 4 generate uplink traffic transferred from the UEs to an application server, and flows 5 through 8 send downlink traffic from an application server to 4 UEs. Finally, flows 9 through 12 send uplink traffic from UEs to an application server. 

We conducted two experiments: in the first experiment, there is only one VM on the type 1 machine. Although there is a separate UPF serving each NS, both UPFs run on the same VM. In the second experiment, there are two VMs set up on the same type 1 machine used for the first experiment and each UPF is supported by a separate dedicated VM. In order to ensure that the aggregate resources allocated to the two NSs remain the same in both experiments, we assigned the same set of cores and non-uniform memory access (NUMA) sockets in both cases. Comparing these two setups allows us to examine how sharing a VM between the two NSs in the first experiment influences the service interference between them. 

\subsubsection{Scenario 1 -- Static Traffic}

\begin{figure}[t]
\centerline{
\includegraphics[width=3.4in]{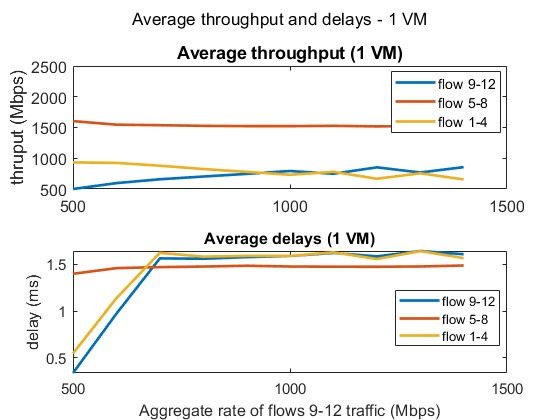}
}
\centerline{(a)}
\centerline{
\includegraphics[width=3.4in]{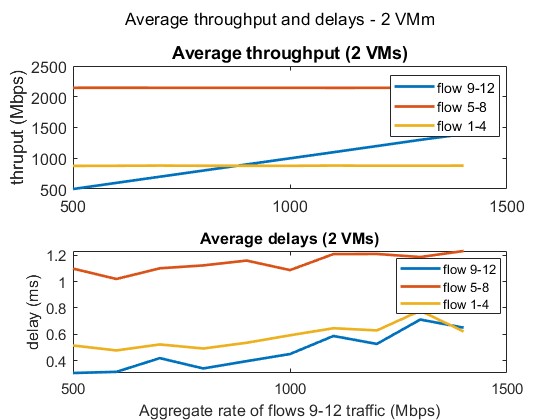}
}
\centerline{(b)}
\caption{Average delays and packet loss rates with constant traffic. (a) single VM, (b) two VMs.}
\label{fig:testbed-0}
\end{figure}

We first consider a scenario where all 12 flows send traffic at a constant bit rate (CBR). However, we change the aggregate traffic rate of flows 9 through 12 and examine how the increasing traffic load handled by NS 2 affects the performance of the 12 flows in the two experiments. 

We use slightly different configurations for the two experiments because we found that the network can handle more traffic when each NS is supported by a dedicated VM (i.e., the second experiment). For this reason, in the first experiment (resp. second experiment), the aggregate traffic of flows 1 through 4 is fixed at 943.7 Mbps (resp. 885 Mbps), and that of flows 5 through 8 is 1.862 Gbps (resp. 2.147 Gbps). Hence, the total traffic through NS 1 handling flows 1 through 8 is higher in the second experiment. The total rate of flows 9 through 12 is varied (approximately) from 500 Mbps to 1.4 Gbps. 

Fig.~\ref{fig:testbed-0} shows the average throughput and delays for different groups of flows as we increase the total rate of flows 9 through 12. The plots show clear service interference between NSs 1 and 2 in the first experiment: as the total rate of flows 9 through 12 increases, the throughput of flows 5 through 8 decreases noticeably, whereas the delays of flows 1 through 4 rise significantly between 500 Mbps and 700 Mbps. On the other hand, in the second experiment the aggregate throughput of flows 1 through 8 remains stable. However, there is a slight elevation in their delays with an increasing rate of flows 9 through 12.

\subsubsection{Scenario 2 -- Time-Varying Traffic}

In the second scenario, while we fix the rates of flows 1 through 11, we vary that of flow 12 over time. The rate of flow 12 changes between two values and is held constant for a fixed duration. The primary goal of the second scenario is to investigate how the varying traffic load on NS 2 affects the performance of the flows handled by both NSs.

\begin{figure}[t]
\centerline{
\includegraphics[width=3.4in]{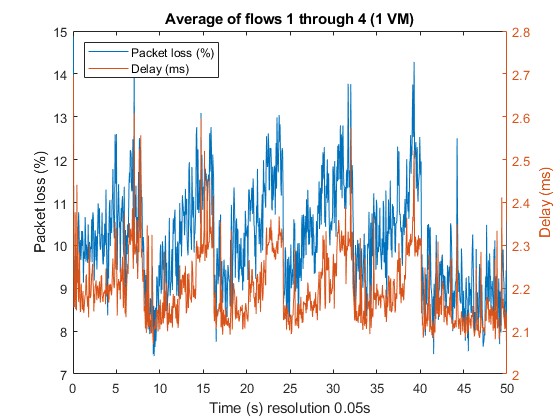}
}
\centerline{
\includegraphics[width=3.4in]{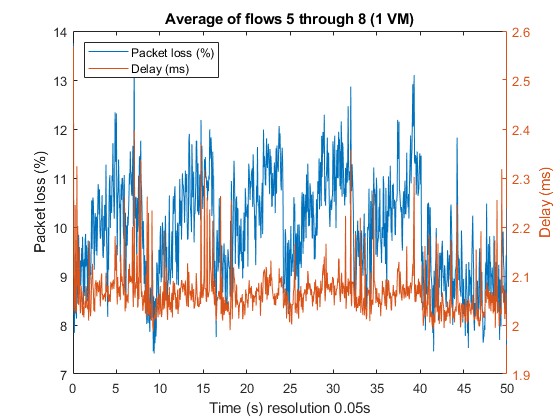}
}
\centerline{
\includegraphics[width=3.4in]{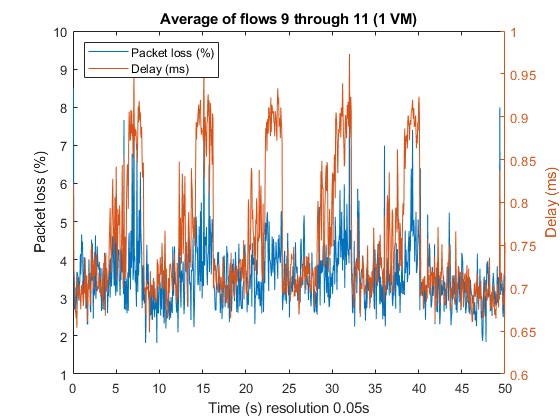}
}
\caption{Average delays and packet loss rates with a single
VM supporting both NSs}
\label{fig:testbed-1}
\end{figure}

Fig.~\ref{fig:testbed-1} plots the average delays and packet loss rates experienced by different groups of flows in the first experiment. The reported numbers are the average of 6 runs. There is a clear pattern in both throughput and delays, which is caused by the time-varying traffic of flow 12; in addition to flows 9 through 11, which share the same NS, flows 1 through 8 supported by a different NS also exhibit a very similar pattern. This suggests that when the NSs are run on the same VM, congestion in one NS may adversely affect the performance of other NSs running on the same VM. 

\begin{figure}[h]
\centerline{
\includegraphics[width=3.4in]{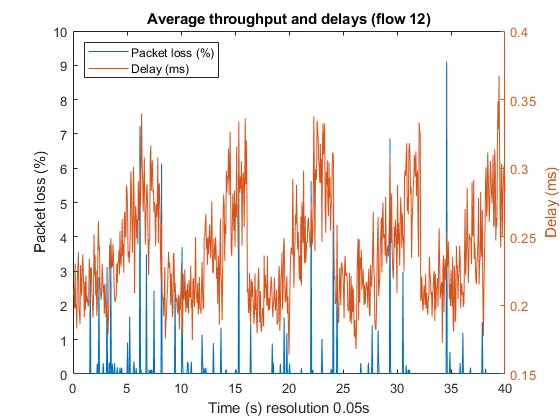}
}
\caption{Average delays and packet loss rates with a dedicated
    VM supporting each NS (flow 12).}
\label{fig:flow12}
\end{figure}

\begin{figure}[h]
\centerline{
\includegraphics[width=3.4in]{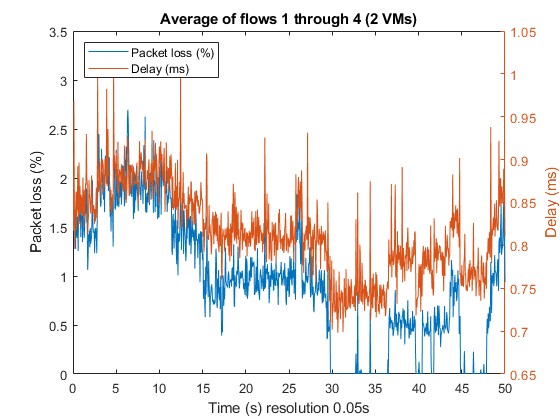}
}
\centerline{
\includegraphics[width=3.4in]{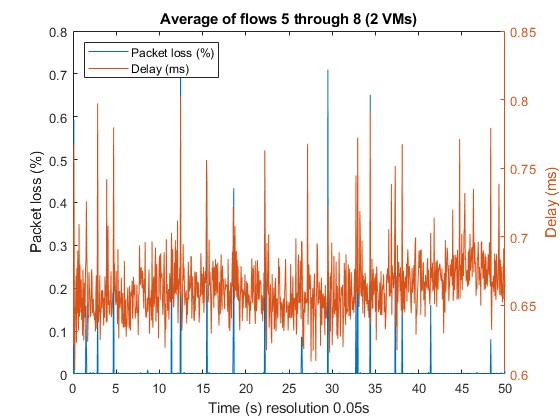}
}
\centerline{
\includegraphics[width=3.4in]{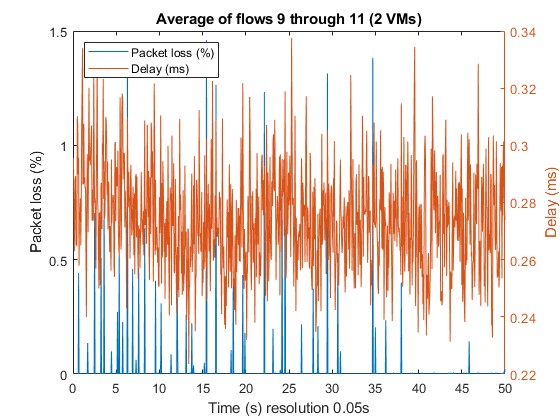}
}
\caption{Average delays and packet loss rates with a dedicated
    VM supporting each NS (flows 1 through 11).}
\label{fig:testbed-2}
\end{figure}

We plot in Figs.~\ref{fig:flow12} and \ref{fig:testbed-2} the average delays and packet loss rates from the second experiment (the average of 6 runs). We plot those of flow 12 separately as it tends to experience somewhat different packet loss rates and delays from other flows as its rate is time-varying. First, we note that flow 12 experiences higher packet losses, which are significant at times, than flows 9 through 11. Furthermore, the pattern exhibited by its delays is consistent with its time-varying traffic. This suggests that NS 2 experiences occasional congestion due to high traffic from flow 12.  

Second, while the time-varying traffic of flow 12 affects the performance of flows 1 through 8 in Fig.~\ref{fig:testbed-2}, the extent of service interference is much smaller than in the first experiment. For example, the packet loss rates of flows 1 through 4 are somewhat higher when NS 2 is congested and flow 12 suffers packet losses in the first 15 seconds of the experiment. However, the service interference is not significant enough to cause their packet loss rates or delays to follow the same pattern shown by flow 12 even when NS 2 experiences congestion. In other words, even though flows 1 through 4 do experience non-negligible packet losses in the first 15 seconds of the experiment, the time-varying traffic of flow 12 does not cause any noticeable pattern in their packet losses or delays that is consistent with that of flow 12. Therefore, the plots indicate that when each NS is supported by a dedicated VM, their interference is reduced in the scenario under consideration.

\section{Proposed Algorithm: Phase 1 - Detection of Shared Resources}	
	\label{sec:Proposed-shared-resources}

In this section, we describe our approach and proposed algorithm for identifying resources shared among NSs. To facilitate our discussion, we focus on a single KPI. When the measurements for multiple KPIs are available, the information can be merged in different ways. For example, the measurements could be used simultaneously or sequentially (i.e., one set of KPI measurements is used at each iteration). 

Let ${\bf M}$ be a $T \times N$ matrix containing the KPI measurements from $N$ NSs over $T$ measurement periods: the $k$-th row of ${\bf M}$ is a vector with the KPI measurements from the $N$ NSs during the $k$-th measurement period and is denoted by ${\bf M}_k$. For simplicity, we assume that there are no missing measurements or rows with missing measurements could be removed. 

The proposed algorithm has three stages: in the first stage, it constructs an {\em undirected interference graph} among NSs based on the strength of pairwise correlations computed using the measurements. The second stage generates {\em a list of maximal cliques} in the interference graph. The third and final stage produces, for each maximal clique in the interference graph, a list of subsets of NSs in the maximal clique, where each subset corresponds to a set of NSs that may share a resource. 

We point out that our focus is on identifying the resources that experience some level of congestion at least occasionally, which causes variations in the KPI measurements of the NSs that utilize them. The resources that are underutilized and do not cause any variations in the KPI measurements (hence no service interference) are of little interest and will not be detected by our proposed algorithm. 
\\ \vspace{-0.1in}

\noindent $\bullet$ {\bf Stage 1. Construction of interference graph --} The purpose of the first stage is to determine {\em pairwise} correlations between two NSs on the basis of the correlations in their KPI measurements. Such pairwise correlations between two NSs are used to determine if they share one or more resources. Based on the empirical correlations, we generate an $N \times N$ 0-1 matrix ${\bf G}$ to capture pairwise correlations: $G_{i,j} = 1$ indicates that NSs $i$ and $j$ display sufficient correlations and $G_{i,j} = 0$ otherwise. 

To this end, we first need to select a suitable measure of correlations or similarity. Although there are several notions of correlations, our numerical studies suggest that the Spearman's rank correlation coefficients work well for our purpose (see Section~\ref{subsec:CorrCoeff} for more details). 

Let $C_{i,j}$ be the Spearman's rank correlation coefficients between NSs $i$ and $j$, and define ${\bf C} := [C_{i,j} : i,j \in \mathcal{S}]$ to be the Spearman's rank correlation coefficient matrix. Because the calculation of each coefficient requires sorting the measurements to determine the ranks, computing $\mathbf{C}$ takes $O(N^2 T\log T)$ time. 

Since the correlation coefficients, as a measure of the degree of correlations between two NSs, tell us how strongly the measurements of two NSs are correlated, in theory we could use a threshold on $C_{i,j}$ to determine which pairs of NSs interfere with each other. Unfortunately, we do not have any prior knowledge of how strong or weak interference and hence correlations could be among the NSs that share resources. For this reason, it is difficult to pre-select a threshold on correlation coefficients for determining pairwise interference among NSs. 

In order to cope with the issue, we propose a clustering-based approach: we use a clustering algorithm to partition the correlation coefficients $C_{i,j}, i \neq j$, into two clusters $\mathcal{C}_0$ and $\mathcal{C}_1$, where the values in $\mathcal{C}_1$ are larger than those in $\mathcal{C}_0$ -- If $C_{i,j} \in \mathcal{C}_1$, we declare that NSs $i$ and $j$ interfere with each other. Otherwise, we assume that they do not. For clustering the correlation coefficients, any reasonable clustering algorithm can be used. For our numerical studies reported in Section~\ref{sec:Numerical}, we use the $k$-means clustering algorithm.

Using the output of the clustering algorithm, we construct an $N \times N$ 0-1 matrix ${\bf G}$, where $G_{i,j} = 1$ if $C_{i,j} \in \mathcal{C}_1$ and $G_{i,j} = 0$ if $C_{i,j} \in \mathcal{C}_0$. We assume $G_{i,i} = 0$ for all $ i \in \{1, \ldots, N\}$. This matrix ${\bf G}$ is the adjacency matrix of the undirected interference graph that will be used in the following stages to identify a list of shared resources and the respective sharing NSs.  
\\ \vspace{-0.1in}

\noindent $\bullet$ {\bf Stage 2. Construction of a list of maximal cliques in the interference graph --} Once the interference graph is constructed in the first stage, we generate a list of all maximal cliques of the interference graph in the second stage. A maximal clique of the interference graph is a complete subgraph, i.e., every pair of NSs have an edge between them in the interference graph, such that if we add any other NS to the subgraph, it is no longer complete. Hence, it represents a largest set of NSs with pairwise interference between every pair of NSs. Note that finding all maximal cliques usually takes exponential time in general, but efficient algorithms are available for large sparse graphs; e.g., \cite{eppstein2013listing} shows that this can be done in $O(Nd3^{d/3})$ where $d$ is the degeneracy number of the graph defined as the smallest number such that every subgraph contains a node of degree at most $d$. Degeneracy is a measure of sparsity, which we believe to be small for the interference graph. 

In general, there could be more than one maximal clique of the interference graph. We denote the list of maximal cliques of the interference graph by $\mathcal{MC} = \{CL_1, CL_2, \ldots, CL_K\}$, where $K$ is the number of maximal cliques.
\\ \vspace{-0.1in}

\noindent $\bullet$ {\bf Stage 3. Identification of a list of shared resources and sharing NSs --} As mentioned earlier, the goal of the last stage is to identify for each maximal clique produced in Stage 2 a set of resources that are shared by distinct subsets of NSs in the maximal clique. We use two examples to illustrate the main challenges to this task. First, consider the example in Fig.~\ref{fig:example} (in Section~\ref{sec:Setup}). In this example, even though there are two shared resources, there is only one maximal clique in the true interference graph which consists of all three NSs since every NS can interfere with the other two NSs via resource 1. Thus, if we map only the maximal clique to a shared resource, we will incorrectly conclude that there is only one resource shared by all 3 NSs. 

\begin{figure}[h]
\centerline{
    \includegraphics[width=1.5in]{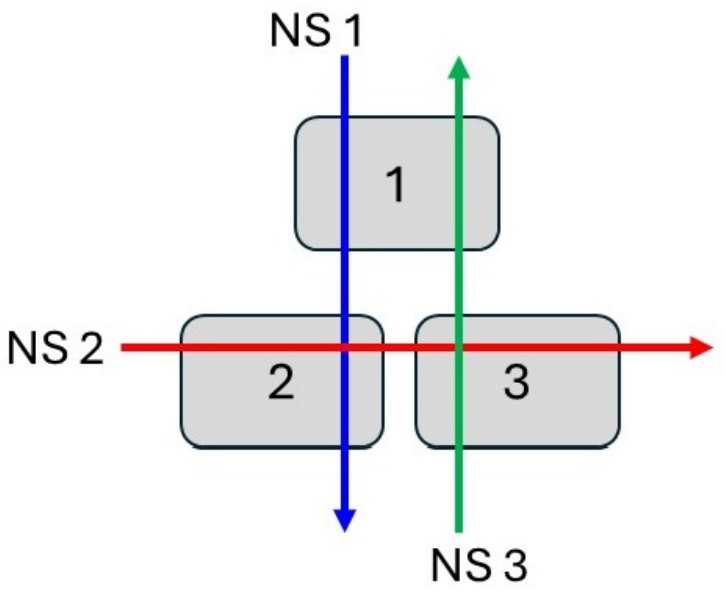}
}
\caption{Example of 3 NSs sharing 3 resources.}
\label{fig:example2}
\end{figure}

Consider another example shown in Fig.~\ref{fig:example2}, where 3 NSs share 3 resources. The unique maximal clique in the interference graph contains all 3 NSs because there is an edge between every pair of NSs. However, unlike in the first example, the maximal clique itself does not correspond to any shared resource; instead, we need to discover 3 shared resources with distinct pairs of sharing NSs, namely three cliques \{NS1, NS2\}, \{NS2, NS3\}, and \{NS1, NS3\}, from the maximal clique \{NS1, NS2, NS3\}. As illustrated by these examples, a key challenge is that, given a maximal clique, we need to correctly identify potentially multiple resources that are shared by different subsets of the maximal clique, using only E2E measurements.

As explained below, for each maximal clique $c$ in $\mathcal{MC}$, the proposed algorithm produces $S_c \subset 2^{c}$, where each subset of $c$ in $S_c$ represents a set of more than one NS that share a resource. Hence, $| \cup_{c \in \mathcal{MC}} S_c |$ is the estimated number of resources that our algorithm determines are shared by distinct sets of NSs.\footnote{Note that it is possible that the same set of NSs may be identified by the algorithm starting with two distinct maximal cliques. However, it will be counted only once in the union $\cup_{c \in \mathcal{MC}} S_c$.} 

Our algorithm is based on FA~\cite{BarthKnott} summarized in Section~\ref{subsec:FA}. We take the position that there are two main sources to the fluctuations in the KPI measurements: (a) changes in the state of the shared resources, and (b) variations in the congestion level within each individual NS. The manner in which the change in the state of a shared resource affects the KPI measurements of sharing NSs is captured by a factor loading associated with the shared resource, and its common factor depends on its state, e.g., the congestion level at a shared resource which influences the measurements. Thus, the latent variables, namely common factors, reflect the state of shared resources in the network. Here we discuss two example KPIs. 

{\bf Example 1. E2E delays --} E2E delays experienced by the NSs during a measurement period can be approximated as the sum of the fixed propagation delays and queueing$+$service delays at the resources. 
When temporary congestion at a shared resource, say a VNF, is caused by one or a few misbehaving NSs, the congestion will likely impact the E2E delays of NSs sharing the VNF to varying degrees. The factor loading associated with the VNF models how the congestion level at the VNF affects the delays experienced by different NSs at the VNF, and the elevation in measured E2E delays of the affected NSs will be determined by an increase in the common factor associated with the VNF (via the corresponding factor loading) and specific factors: \\
First, the increase in the E2E delays for all affected NSs will be captured by $f_r {\bf L}_r$ for some $r$ (corresponding to the VNF in \eqref{eq:FA1}), where ${\bf L}_r$ is the factor loading associated with the VNF and $f_r$ is the corresponding common factor; the more congested the VNF is, the larger the common factor $f_r$ will be, thereby increasing the delays experienced at the VNF. We point out that because the elements of the factor loading ${\bf L}_r$ can vary, congestion at the VNF can affect the delays of affected NSs differently. 

Second, assuming proper resource provisioning, even when one or a few NSs misbehave and cause congestion, other well-behaved NSs that adhere to their service level agreements should not experience a significant/unacceptable increase in delays. This suggests that most of the increase in the delays experienced by well-behaved NSs should come from the elevation in the common factor $f_r$ (via factor loading ${\bf L}_r$) and their specific factors should be small. On the other hand, the misbehaving NSs should suffer greater delays due to the internal congestion within the NS, which will be captured by larger specific factors (in addition to the elevation in the common factor). Therefore, the factor loading models how the congestion level at a shared resource affects the delays experienced by all sharing NSs, whereas specific factors model the larger delays suffered by congested/misbehaving NSs. 

{\bf Example 2. E2E packet loss rates --} Denote the packet loss rate of NS $i$ at a resource $j$ by $p_{i,j}$, and suppose that packet loss rates at shared resources are small, i.e., $p_{i,j} \ll 1$. In this case, under a suitable independence assumption, the E2E packet loss rate of NS $i$ is given by $1 - \prod_{j \in \mathcal{R}^i} (1 - p_{i,j}) \approx \sum_{j \in \mathcal{R}^i} p_{i,j}$, where $\mathcal{R}^i$ is the set of resources utilized by NS $i$. This suggests that the E2E packet loss rate is approximately additive when they are small. Therefore, if the overall congestion level of a shared resource affects the packet loss rates of different NSs at the resource in a similar fashion (not including additional packet loss rates they may experience due to respective internal congestion), the FA can be employed to model the packet loss rates of NSs.  
\\ \vspace{-0.1in}

Let us continue with the discussion on how FA is used in our algorithm. For each maximal clique $c \in \mathcal{MC}$, let ${\bf M}^c$ be the submatrix of ${\bf M}$ consisting of 
only the columns of ${\bf M}$ corresponding to the NSs in $c$. We perform FA on ${\bf M}^c$ and identify a suitable factor loading matrix ${\bf L}^c$ so that the rows of ${\bf M}^c$ minus the mean vector can be written as linear combinations of the rows of ${\bf L}^c$ (i.e., factor loadings) plus specific factors as follows.   
\beqa
{\bf M}^c_k = \sum_{r=1}^{q^c} f^c_{k,r} {\bf L}^c_r + \boldsymbol{\epsilon}^c_k  + \overline{\bf m}^c, \quad k = 1, 2, \ldots, T, 
    \label{eq:M_c}
\eeqa
where the weights $f^c_{k,r}$ are the common factors, $\bepsilon^c_k$ contains the specific factors, and $\overline{\bf m}^c$ is the mean measurement vector for NSs in $c$.\footnote{Recall that, for symmetrized meeasurements, $\overline{\bf m}^c = {\bf 0}$.}  In a matrix form, we have the following equation: 
\beqan
{\bf M}^c = {\bf F}^c {\bf L}^c + {\bf E}^c + \overline{\bf M}^c
\eeqan
where ${\bf F}^c$ is a $T \times q^c$ matrix with common factors, ${\bf L}^c$ is a $q^c \times |c|$ matrix with loading factors, ${\bf E}^c$ is a $T \times |c|$ matrix containing specific factors (with $\bepsilon^c_k$ as the $k$-th row), and $\overline{\bf M}^c$ is a $T \times |c|$ matrix whose rows are all equal to $\overline{\bf m}^c$.




In the proposed algorithm, for each maximal clique $c \in \mathcal{MC}$, we vary the value of $q^c$ to find a suitable number of factor loadings to consider and, for each fixed $q^c$, we employ the maximum likelihood estimation to determine the factor loadings in ${\bf L}^c$ and the variances of specific factors. Then, we choose the value of $q^c$ that yields the largest log likelihood. Note that this step can be done efficiently using existing nonlinear optimization algorithms such as those in \cite{dempster1977maximum, zhao2008ml} which often run in  $O(T|c|^2+K|c|^3)$ time for $K$ iterations. Note also that $|c|\le d+1$ for any $c \in \mathcal{MC}$, where $d$ is the degeneracy number of the interference graph.

$\bullet$ {\bf Computation of common factors: } 
In order to determine which NSs misbehave and cause adverse interference to other NSs using the algorithm described in the next section, we need the common factors ${\bf F} = [f_{k,i} : k = 1, \ldots, T; \ i = 1, \ldots, N]$. For our algorithm, while we adopt the factor loadings from the FA-based algorithm, we do not use its common factors. The reason for this is that the common factors computed above are calculated separately in isolation for each maximal clique. As a result, when the common factors for the shared resources associated with a maximal clique are computed, they are based on the measurements that include the contributions of other shared resources utilized by (some of) the NSs in the maximal cliques. To mitigate this effect, we recompute all common factors simultaneously based on the global information after all factor loadings are computed.  

We formulate the problem of estimating ${\bf F}$ as least squares problems: let $\tilde{\bf M}$ be the centered measurement matrix with $\tilde{\bf M}_k := {\bf M}_k - \overline{\bf m}$, $k = 1, \ldots, T$. Note that for symmetrized measurements, $\tilde{\bf M} = {\bf M}$. For each $k = 1, \ldots, T$, the common factor vector ${\bf F}_k$ is obtained as a solution to the following optimization.   
\beqa
\mbox{minimize}_{\check{\bf f} \in \R^q} &  
    \| \check{\bf f} \ {\bf L} - \tilde{\bf M}_k \|_2^2 
    \label{eq:LSP}
\eeqa
The solution to \eqref{eq:LSP} is given by ${\bf f}_k^* = \tilde{\bf M}_k {\bf L}^T ({\bf L \ L}^T)^{-1}$, $k = 1, \ldots, T$. In a matrix form, 
\beqa
{\bf F}^* = \tilde{\bf M} {\bf L}^T ({\bf L \ L}^T)^{-1} \ . 
    \label{eq:CommonFactor}
\eeqa

In the remainder of the paper, we will use the following notation: assume that $q$ is the total number of factor loadings, i.e., estimated number of shared resources. 
\bitem

\item ${\bf L}$ - $q \times N$ matrix that contains the $q$ estimated factor loadings as its rows. For each factor loading obtained from considering a maximal clique $c$, to construct an $N$-dimensional row vector we add zeros for the NSs not in $c$ to indicate they do not share resources identified from the maximal clique $c$. We normalize the factor loadings by the largest element so that $\max_{i = 1, \ldots, N} l_{j,i} = 1$ for all $j = 1, \ldots, q$; 

\item ${\bf F}$ - $T \times q$ matrix with common factors with the $k$-th row containing the common factors of the $q$ factor loadings for the $k$-th measurements ${\bf M}_k$. As explained above, ${\bf F}$ is obtained using \eqref{eq:CommonFactor};

\item ${\bf E}$ - $T \times N$ matrix with the specific factors associated with ${\bf M}_k$ in the $k$-th row. We set ${\bf E} = \tilde{\bf M} - {\bf F} {\bf L}$; and 

\item $\overline{\bf M}$ - $T \times N$ matrix whose rows are all equal to the
mean vector $\overline{\bf m}$ (which will be a zero vector for symmetrized measurements)

\eitem

Before we proceed, we explain why we adopt FA, but not PCA. Suppose that the state of a shared resource affects all sharing NSs in a qualitatively similar manner, i.e., the elements for the affected NSs have the same sign. Since we wish to approximate the measurements using a linear model that captures the effects of each shared resource on KPI measurements using a vector, it is natural to allow two vectors with common NSs to be non-orthogonal when one or more NSs utilize both resources. Consider the example shown in Fig.~\ref{fig:example}.  The factor loadings associated with the two shared resources 1 and 2 will not be orthogonal because both resources are shared by NSs 1 and 2. Similarly, the factor loadings associated with the three shared resources in the second example in Fig.~\ref{fig:example2} will not be orthogonal.

\section{Proposed Algorithm: Phase 2 - Detection of Misbehaving Network Slices}
    \label{sec:proposed-misbehaving-NSs}

In the previous section, we outlined the algorithm for detecting shared resources in the network. In this section, we describe how the output of the algorithm can be utilized to identify misbehaving NSs that may cause potentially harmful service interference to other NSs. Together with the output from the first algorithm, this allows us to isolate the NSs affected by the misbehaving NSs via shared resources. 

From the discussion at the end of the previous section, we have 
\beqa
{\bf F} {\bf L} + {\bf E} = \tilde{\bf M} = ({\bf M} - \overline{\bf M}) \ . 
    \label{eq:model1}
\eeqa
Eq.~\eqref{eq:model1} tells us that the difference $M_{k,i} - \overline{m}_{i}$, which models the deviation in the KPI measurement from its mean, can be partitioned into two components -- one determined by the aggregate utilization levels of shared resources which affect the KPI measurements of all affected NSs, and the other dependent only on own state or utilization level in relation to the resources allocated to the NS: \\
First, the overall congestion or utilization level of a shared resource (e.g., a physical device/machine supporting multiple VMs or VNFs) affects the traffic and hence KPI measurements of all NSs that share the resource. As explained earlier, this dependence of KPI measurements on the overall utilization levels of resources is captured via the factor loadings and common factors in the first term on the left-hand side (LHS) of \eqref{eq:model1}. Moreover, the qualitative impact of time-varying utilization level of a shared resource on KPI measurements of affected NSs is likely similar and positively correlated, even though quantitative effects may vary from one affected NS to another (which the associated factor loadings aim to capture).

Since the dependence of NSs' KPI measurements on the overall congestion levels at shared resources is modeled by the first term, the remaining fluctuations in KPI measurements can be attributed to the traffic dynamics within each individual NS plus measurement noise and are expected to be independent of each other. Specific factors model this impact of the individual NSs on their own KPI measurements. They depend on several factors, including the resources allocated to individual NSs in accordance with service level agreements. This suggests that specific factors provide information about relative congestion levels of NSs compared with the congestion levels of the resources they utilize; NSs with higher congestion\textit{} will likely have larger positive specific factors, while those with low utilization levels will likely have smaller, possibly negative specific factors. Our algorithm exploits this observation.

\subsection{Determination of Misbehaving Network Slices}
    \label{subsec:MisbehavingNS}

The intuition behind our proposed approach is as follows: in order for a misbehaving NS to bring about interference to other NSs, it needs to trigger congestion at one or more shared resources, which in turn likely causes both the common factors associated with the congested resources and its own specific factor to change significantly. On the other hand, NSs that share congested resources with misbehaving NSs may experience elevated KPI measurements due to congestion at shared resources. However, these deviations in KPI measurements will be captured mostly by the common factors as explained earlier, and for well-behaved NSs, their specific factors should remain small. This suggests that, in order to find NSs that cause undesirable interference to other NSs, we should look for NSs that not only have larger specific factors, but also utilize one or more congested resources at the same time. Our algorithm leverages this intuition. 

First, we identify the set of shared resources that experienced some congestion by considering their common factors: choose $\gamma \geq 0$, which is a threshold on common factors, and define $\hat{f}_{k, l} = f_{k,l} \cdot \indicate{f_{k, l} \geq \gamma}$, $k = 1, \ldots, T$, and $l = 1, \ldots, q$, and let $\hat{\bf F} = \big[ \hat{f}_{k,l} : k = 1, \ldots, T; l = 1, \ldots, q \big]$. The interpretation is that if $\hat{f}_{k,l} > 0$, resource $l$ contributes to the KPI measurements of NSs utilizing it during measurement period $k$, and the choice of $\gamma$ determines how high the congestion at a shared resource ought to be in order to be considered. For instance, if $\gamma = 0$, our algorithm removes the negative common factors in our algorithm. Note that some common factors may be negative depending on the mean vector $\overline{\bf m}$ and measurement noise. 

Second, we compute for each NS the aggregate effects of the common factors of utilized resources, which we call the {\em aggregate common factors}, by multiplying $\hat{\bf F}$ by ${\bf L}$. Note that the difference between the first term on the LHS of \eqref{eq:model1} and $\hat{\bf F} {\bf L} =: {\bf W}$ is that the latter tries to remove the noise in common factors coming from resources whose common factors stay below the threshold $\gamma$. Recall that the factor loadings were normalized so that the maximum element of each factor loading is equal to one. The reason for this is that we want the KPI measurements and the common factors to have the same unit for the NS that is most affected by each resource. In other words, as the common factor changes, it changes the KPI of measurements of the most affected NS(s) by the same amount.  

Third, to calculate specific factors in ${\bf E}$, we substitute $\hat{\bf F}$ in place of ${\bf F}$ in \eqref{eq:model1} and subtract $\hat{\bf F} {\bf L}$ from $\tilde{\bf M}$ to obtain $\hat{\bf E} = \tilde{\bf M} - \hat{\bf F} {\bf L}$. Then, we compute the weighted specific factors, which are the specific factors multiplied by the aggregate common factors, given by ${\bf W} \odot \hat{\bf E} =: {\bf \Theta}$, where $\odot$ denotes the Hadamard product. 

Finally, we compute the total weighted specific factors of the NSs: 
\[
\Gamma_i := \sum_{k=1}^T \max\big( 0, {\theta}_{k,i} \big) , 
    \quad i = 1, \ldots, N . 
\]
Then, we partition $\{ \Gamma_i : i = 1, \ldots, n \}$ into two clusters using a clustering algorithm. For our numerical studies in Section~\ref{sec:Numerical}, we use $k$-means clustering algorithm with $k = 2$. Our algorithm returns the set of NSs in the cluster with larger values of $\Gamma_i$ as potentially misbehaving NSs. 

We can modify the algorithm so that an optimal number of clusters is identified using some criterion (e.g., silhouette value) instead of fixing the number of clusters. This will allow us to consider the distribution of the total weighted specific factors among different clusters to acquire additional information and determine a suitable threshold on the total weighted specific factors.

\section{Numerical Studies}
	\label{sec:Numerical}

In order to evaluate the performance of the proposed algorithms, we conducted numerical studies using Matlab. We first describe the setup for carrying out the simulation, followed by our findings.\footnote{Even though a testbed was used to demonstrate the service interference between NSs earlier, the testbed does not scale to a suitable size. For this reason, we use a simulation rather than the testbed results.}


\subsection{Simulation Setup}	\label{subsec:SimSetup}

For the simulation, we consider 15 resources shared by 50 NSs. Although they are not included here due to a space constraint, additional studies were performed with 20 NSs sharing 6 resources to examine the scalability and accuracy of the proposed algorithm with varying network size. These results can be found in~\cite{ArXiV}.\footnote{Recall that we are only interested in resources that experience varying levels of congestion during the monitoring period, and other resources that do not experience congestion during the period need not be modeled. Hence, the resources modeled in the simulation should be viewed as those whose time-varying congestion level affects the measurements.} We denote the number of shared resources by $R$. As mentioned earlier, these resources could be shared communication links, VNFs or other physical resources, including CPUs and memory. 

For each shared resource $j \in \mathcal{R} := \{1, 2, \ldots, R\}$, let $\mathcal{N}_j$ be the set of NSs that utilize resource $j$. Similarly, for each $i \in \mathcal{N} := \{1, 2, \ldots, N\}$, $\mathcal{R}^i$ denotes the set of resources utilized by NS $i$. This is captured by an assignment matrix ${\bf A}$, which is an $R \times N$ 0-1 matrix: when $A_{j,i} = 1$, NS $i$ utilizes and shares resource $j$, and if $A_{j,i} = 0$, NS $i$ does not utilize resource $j$. 

In our simulation studies, we focus on one KPI measurements, namely E2E delays. The delays experienced by the traffic belonging to an NS at a resource depends on two factors -- (a) the total utilization level of the resource and (b) the individual NS utilization level. We vary the relative weights given to these two factors to study how the strength of the interference among the NSs at the shared resources affects the performance of the proposed algorithm; as more weight is assigned to the latter, the interference among the NSs sharing resources diminishes. 

\noindent $\bullet$ {\bf Dynamics of NS utilization levels:} The utilization level of each NS during measurement period $k$ is determined by a Markov chain (MC) with 4 states, and the MCs for different NSs are independent. The transition matrix for each MC is selected randomly at the beginning of each run; we first generate a 4$\times$4 random matrix, where off-diagonal elements are i.i.d. Uniform(0, 1) random variables and diagonal elements are all zero. Then, we normalize each row (without the diagonal element) so that the row sums are all equal to 0.75, sort entries by decreasing value, and then set the diagonal elements to 0.25. The latter implies that the MC stays at the same state with probability 0.25.

The utilization level vector (associated with the MC states) is [0.2 0.5 0.7 0.9]. For example, when the state of the MC for an NS is 2, its utilization level is assumed to be 0.5. Note that we sort the transition probabilities so that the state transitions to lower utilization states with higher probability when it moves out of the current state. This is done to simulate scenarios in which the average utilization of the NSs is relatively low, but they can experience congestion time to time. 

\noindent $\bullet$ {\bf Utilization level of shared resources:} Let $U_i(k)$, $i \in \mathcal{N}$ and $k \in \N := \{1, 2, \ldots \}$, be the utilization level of NS $i$ during measurement period $k$. The utilization level of resource $j$ during measurement period $k$, denoted by $V_j(k)$, is equal to the average utilization of the NSs that share it:\footnote{This implies that the capacity of a shared resource is proportional to the number of sharing NSs.}
\[
V_j(k) = \frac{1}{|\mathcal{N}_j|} \sum_{i \in \mathcal{N}_j} U_i(k), \quad j \in \mathcal{R} 
\] 

\noindent $\bullet$ {\bf E2E delay measurements:} As mentioned earlier, the E2E delay experienced by NS $i$ during a measurement period is determined not only by its own utilization level, but also by the total utilization levels at the resources in $\mathcal{R}^i$. In our simulation, we compute it with the help of two functions, $g: \R_+ \to \R_+$ and $h: \R_+ \to \R_+$: The function $g$ determines the delay at a resource as a function of its utilization level and captures the delay common to all sharing NSs. The function $h$ models the additional delays, which depend on the utilization levels of individual NSs. 

The E2E delays experienced by NS $i$'s traffic during measurement period $k$ are given by 
\beqan
D_i(k)
\myeq \Big( w_S \sum_{j \in \mathcal{R}^i} g\big( V_j(k) \big) 
+ (1 - w_S) h\big( U_i(k) \big) + n_i(k) \Big)_+ 
\eeqan
where $(\cdot)_+ = \max(0, \cdot)$. The summands in the first term (without $w_S$) on the right-hand side (RHS) represent the delays at the resources utilized by NS $i$ due to possible congestion. The second term captures an  additional delay NS $i$ experiences when its traffic load is high. Finally, $n_i(k)$ models the sum of a fixed delay and measurement noise, where the noise is given by the deterministic part of the E2E delay, namely $w_S \sum_{j \in \mathcal{R}^i} g\big( V_j(k) \big) + (1 - w_S) h\big( U_i(k) \big)$, multiplied by a Gaussian random variable with mean 0 and variance $\sigma^2$. Note that $\sigma$ is the standard deviation of noisy delay when the deterministic delay is one. We study the effects of $\sigma^2$ using numerical results. For numerical studies, we use the following functions:
\beqan
g(x) = (x-0.6)_+^2 \ \mbox{ and } \ h(y) = (y - 0.65)_+^2 
\eeqan

The weight $w_S$ is used to vary the strength of interference among NSs sharing the resources; when $w_S$ is small, the delay of an NS mostly depends on its own utilization and the activities of other NSs have little effect, indicating weak interference. As $w_S$ increases, the delays experienced by an NS are affected more by the aggregate traffic loads at the shared resources, modeling stronger interference. We will examine the effects of $w_S$ in Section~\ref{subsec:ResourceDetection}.

\subsection{Numerical Results}	
    \label{subsec:ResourceDetection}

We vary (a) the number of available samples or sample size $T \in \{300, 500, 700, 1000\}$ and (b) the weight $w_S \in \{0.1, 0.15, 0.2, 0.25, 0.3\}$. The measurement noise variance is fixed at $\sigma^2 = 0.1^2$. In subsection~\ref{subsubsec:Robustness}, we study how the noise variance affects the performance of the proposed algorithm for identifying shared resources. Every NS utilizes at least one resource, and each resource is shared by at least two NSs. This guarantees that every NS shares at least one resource with one or more NSs. Furthermore, we ensure that no two resources are shared by the identical set of NSs.\footnote{Since our goal is to 
identify different sets of NSs that interfere with each other, it is not important whether one shared resource introduces the interference among a set of NSs or it is caused by more than one resource shared by the same set of NSs.}

\subsubsection{Effects of weight $w_S$ on correlations in measurements between network slices sharing resources and their standard deviations}
    \label{subsubsec:wS}

Recall that the weight $w_S$ is used to change the strength of interference among the NSs sharing resources. For this reason, We first examine how the correlations in measurements between NSs sharing resources change with $w_S$.

\begin{figure}[h]
\centerline{
\includegraphics[width=1.75in]{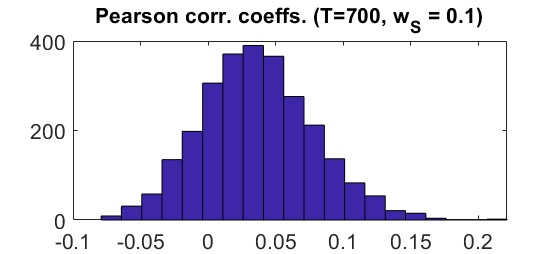}
\includegraphics[width=1.75in]{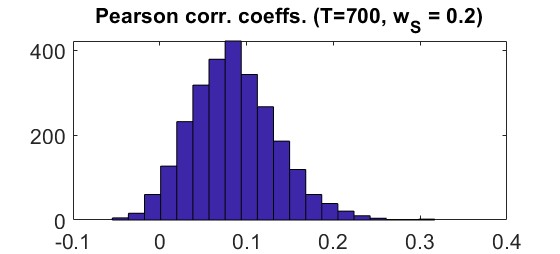}
}
\centerline{
\includegraphics[width=1.75in]{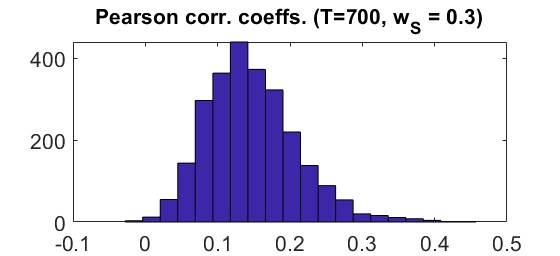}
}
\caption{Pearson correlation coefficients between network slices sharing resources for $w_S =$ 0.1, 0.2, and 0.3 from 25 random runs ($T = 700, \sigma = 0.1$).}
\label{fig:PCC_w010_020}
\end{figure}    

Fig.~\ref{fig:PCC_w010_020} plots the Pearson correlation coefficients (PCCs) for $w_S =$ 0.1, 0.2 and 0.3.\footnote{We use the Pearson correlation coefficients to measure the strenth of correlations because it is commonly used and its values are well understood for measuring correlation strength. For example, PCCs with their absolute values below 0.3 are typically considered weak correlations, whereas absolute values in [0.3, 0.5] are deemed moderate.} It is clear from the plots that the PCCs between NSs sharing resources are small: for $w_S = 0.1$, most coefficients are smaller than 0.1 with both the mean and the median around 0.03, and for $w_S = 0.2$, almost all PCCs are smaller than 0.2 with the median less than 0.1, which suggest weak correlations in the measurements. For $w_S = 0.3$, the PCCs mostly lie between 0.05 and 0.2 with only a few PCCs above 0.25. Thus, small PCCs indicate weak to at most moderate interference between NSs even for $w_S = 0.3$ among the NSs sharing one or more resources. 

\begin{figure}[h]
\centerline{
\includegraphics[width=1.75in]{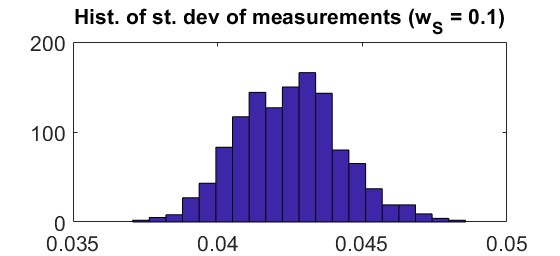}
\includegraphics[width=1.75in]{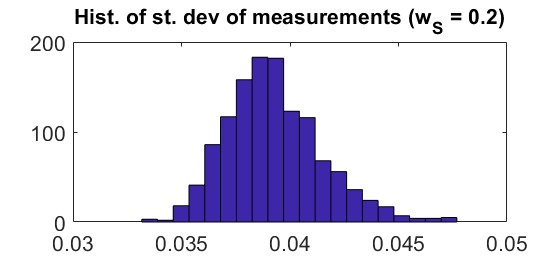}
}
\centerline{
\includegraphics[width=1.75in]{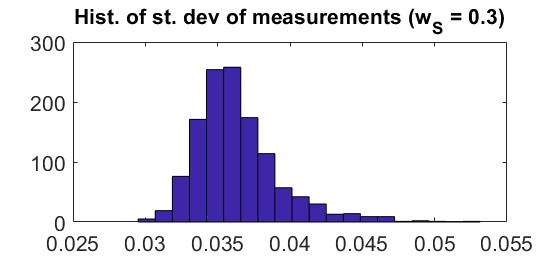}
}
\caption{Histograms of standard deviations of delay measurements for 50 NSs from 25 random runs ($T = 700, \sigma = 0.1$).}
\label{fig:SD}
\end{figure}

In addition to the Pearson correlation coefficients, we also examined the standard deviation of the measurements for each NS. Fig.~\ref{fig:SD} shows the histogram of the standard deviation of the 50 NSs for $T = 700$ and $w_S \in$ \{0.1, 0.2, 0.3\}, which are collected from 20 random runs. The plots reveal that the standard deviation tends to decrease slightly with increasing $w_S$. This is intuitive and expected; as $w_S$ increases, NS delay measurements are more affected by the {\em average} traffic load of resources and less by own traffic load.\footnote{Recall that the capacity of a resource is assumed proportional to the number of sharing NSs.} The average traffic load is the sum of {\em independent} random processes normalized by the number of sharing NSs and, hence, has smaller variance than that of individual traffic load. As a result, increasing $w_S$ reduces the variance in delay measurements.

\subsubsection{Identification of shared resources}

Here we evaluate the performance of our algorithm for identifying shared resources (Section~\ref{sec:Proposed-shared-resources}). For each fixed set of parameters, we report the average of 100 randomly generated resource assignment matrices ${\bf A}$. 

$\bullet$ {\bf Performance metrics:} The first algorithm produces as a part of the output an estimate of the assignment matrix ${\bf A}$, which we denote by $\tilde{\bf A}$. The number of rows in $\tilde{\bf A}$, say $J$, is the estimated number of distinct shared resources, which is equal to $q$, i.e., the number of factor loadings in Section~\ref{sec:Proposed-shared-resources}. We obtain $\tilde{\bf A}$ by applying a threshold $\eta > 0$ to ${\bf L}$. In other words, $\tilde{A}_{j,i} = 1$ if $l_{j,i} \geq \eta$ and $\tilde{A}_{j,i} = 0$ otherwise. For our numerical studies, we select $\eta = 0.15$. However, the performance is not sensitive to the choice of $\eta$ in the studied interval [0.1, 0.2].

The order of the rows in $\tilde{\bf A}$ is arbitrary and is not important. Instead, the positions of the elements equal to 1 in the $j$-th row, $j \in \{1, \ldots, J\}$, tell us which NSs the algorithm believes share some resource. For instance, suppose that the algorithm returns the following matrix as the output for a scenario with 3 NSs.
\beqan
\tilde{\bf A} = \begin{bmatrix}
1 & 1 & 0 \\
1 & 0 & 1 
\end{bmatrix}
\eeqan
In this case, the algorithm determined that there are two shared resources -- one resource is shared by NSs 1 and 2, and the other resource is shared by NSs 1 and 3. 

For each assignment matrix ${\bf A}$, we compute the fraction of correctly identified shared resources as follows.   
\[
\frac{1}{R} \sum_{j = 1}^{R} \left( \sum_{j'=1}^J \indicate{{\bf A}_{j} = \tilde{\bf A}_{j'} } \right)
\]
where ${\bf A}_{j}$ and $\tilde{\bf A}_{j'}$ are the $j$-th row of ${\bf A}$ and the $j'$-th row of $\tilde{\bf A}$, respectively. The plots show the average obtained using 25 random assignment matrices. Note that the expression inside the parentheses is equal to 1 if one of rows in $\tilde{\bf A}$ matches the set of NSs sharing the $j$-th resource and is equal to 0 otherwise. 

In addition to the fraction of correctly identified resources with right sets of sharing NSs, we also consider `coverings': we say that a resource $j$ is {\em covered} by the $j^*$-th row in $\tilde{\bf A}$ if the difference $\tilde{\bf A}_{j^*} - {\bf A}_j$ is a non-negative (row) vector and that $\tilde{\bf A}_{j^*}$ is a {\em covering} for ${\bf A}_j$. Note that when $\tilde{\bf A}_{j^*} = {\bf A}_j$, $\tilde{\bf A}_{j^*}$ is still a covering of ${\bf A}_j$. 

The reason for considering such coverings is two-fold: first, $\tilde{\bf A}_{j^*}$ correctly identifies all sharing NSs, even though it may also mistakenly include one or more NSs that do not share the resource (which we refer to as {\em false positives}). In this sense, it may provide additional useful information when the output of our algorithm is used to take further actions, e.g., migrating NSs that cause significant performance degradation to other NSs. Second, by comparing the fraction of resources for which the output provides a covering to that of correctly identified resources, we can show that most of the coverings are in fact correct identifications without any false positive.

\begin{figure}[h]
\centerline{
\includegraphics[width=1.75in]{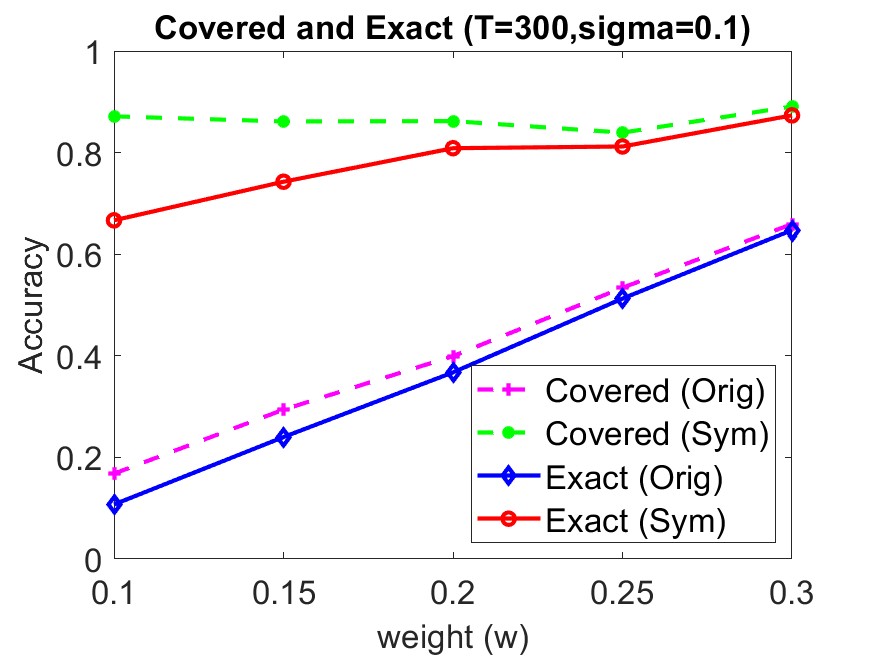}
\includegraphics[width=1.75in]{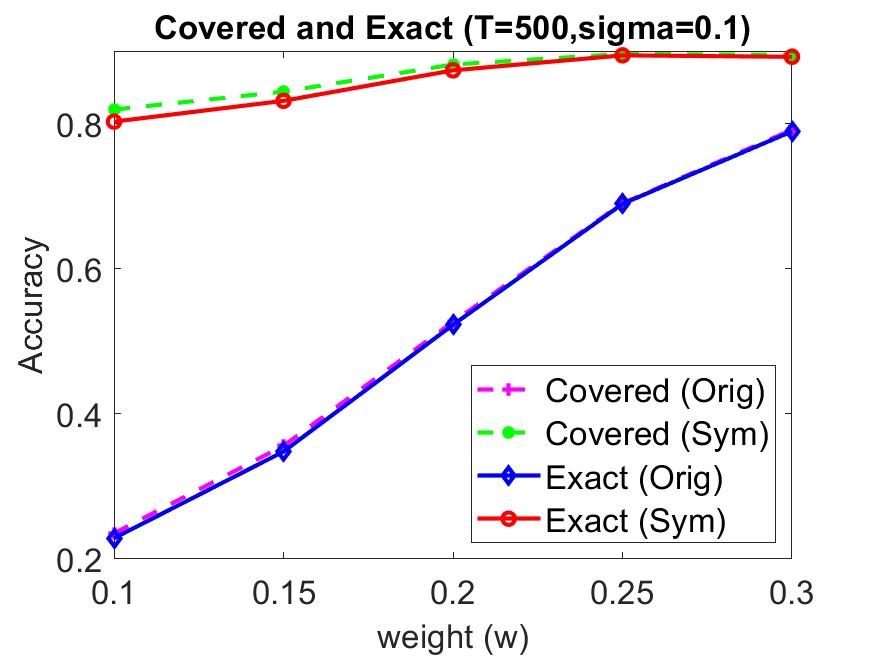}
}
\centerline{
\includegraphics[width=1.75in]{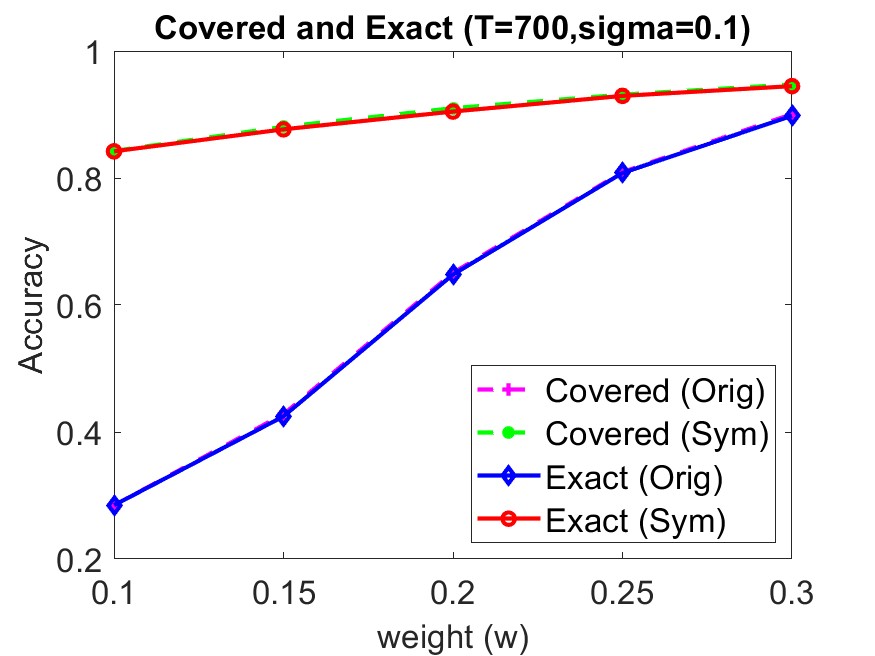}
\includegraphics[width=1.75in]{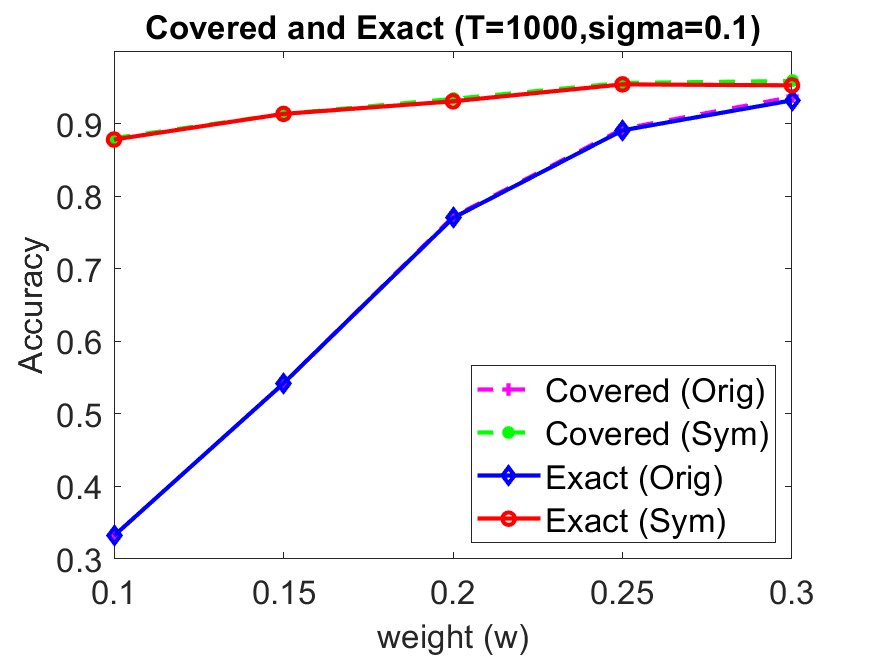}
}
\caption{Accuracy of the proposed algorithm ($\sigma = 0.1$).}
\label{fig:exact-var10}
\end{figure}

Fig.~\ref{fig:exact-var10} plots (i) the fraction of correctly identified resources and respective sharing NSs (\texttt{Exact (Orig)} and \texttt{Exact (Sym)} in the figures) and (ii) the fraction of resources covered by a covering discussed earlier (\texttt{Covered (Orig)} and \texttt{Covered (Sym)}) as a function of the weight $w_S$ for four different sample sizes $T$. The difference between \texttt{(Orig)} and \texttt{(Sym)} is that for \texttt{(Orig)} the original measurements prior to symmetrization are used and \texttt{(Sym)} uses symmetrized measurements discussed in Section~\ref{subsec:SymVsAsym}. The plots show that when either the weight $w_S$ or the number of measurements $T$ is small, using symmetrized measurements is beneficial and leads to a significant improvement in accuracy. However, as both $w_S$ and $T$ become larger, the benefits are only marginal or vanish entirely. This is consistent with our intuition explained earlier. 

It is clear from the plots that, as expected, the accuracy of the algorithm improves with the number of samples $T$. This can be attributed to two main reasons: first, when the sample size $T$ is too small, the provided measurements may not be sufficient to allow the algorithm to see all existing interference patterns among NSs and, as a result, the algorithm fails to identify some shared resources due to insufficient observations. Second, as we will show in the following section, when the sample size is small, pairwise correlation coefficients used to construct the interference graph tend to be noisy with larger variance. Consequently, it becomes difficult to correctly identify pairwise interference, leading to missing edges between NSs in the interference graph produced in Stage 1. This naturally causes the algorithm to miss the corresponding shared resources. 

Second, the plots suggest that, except for when $T = 300$, there is no noticeable difference between the fraction of resources correctly identified and that covered by coverings. For example, for $T = 1,000$ the plots nearly coincide and there is no discernible discrepancy between them. This suggests that our algorithm does not produce many false positives in its output when a sufficient number of measurements are available.

Finally, we note that when symmetrized measurements are used, reasonable accuracy (greater than 80 percent) can be achieved even for relatively small $w_S$ and $T$. For instance, for $w_S = 0.2$ and $T = 300$, the proposed algorithm achieves accuracy of over 80 percent. Recall from Section~\ref{subsec:CorrCoeff} that when $w_S \leq 0.2$, the Pearson correlation coefficients are small (with most being smaller than 0.2) even for $T = 700$, indicating weak correlations in measurements among NSs sharing resources.

\begin{figure}[h]
\centerline{
\includegraphics[width=1.75in]{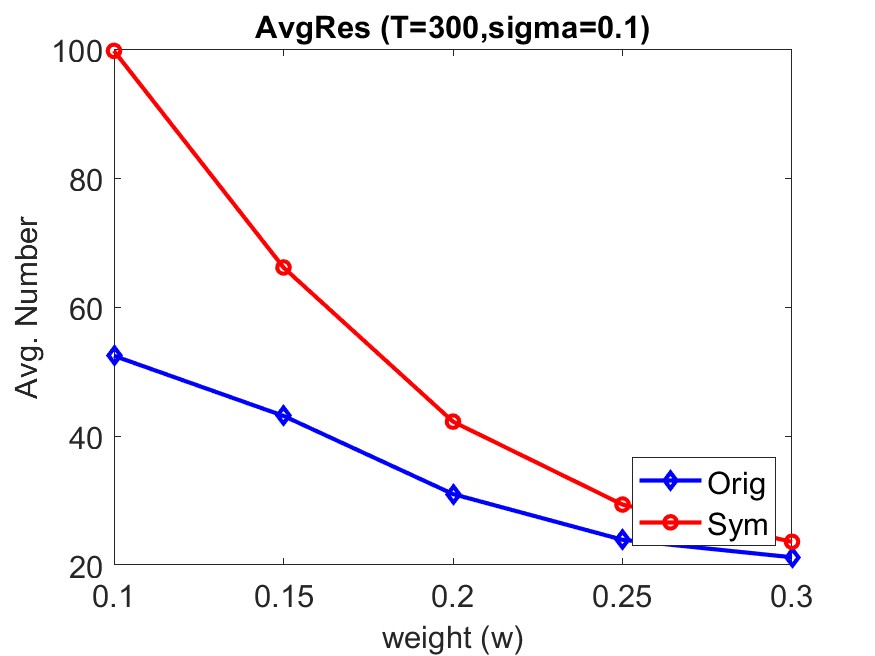}
\includegraphics[width=1.75in]{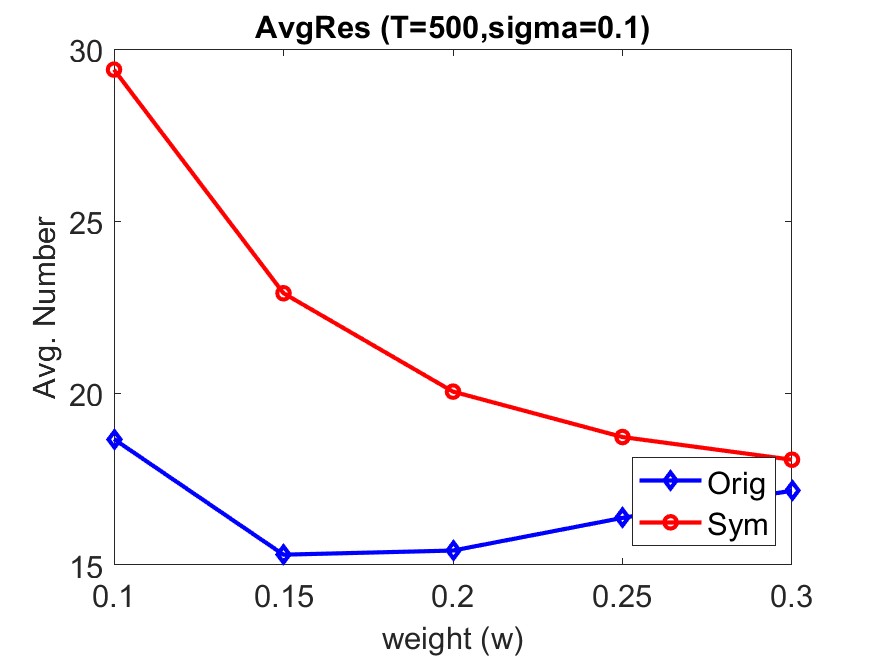}
}
\centerline{
\includegraphics[width=1.75in]{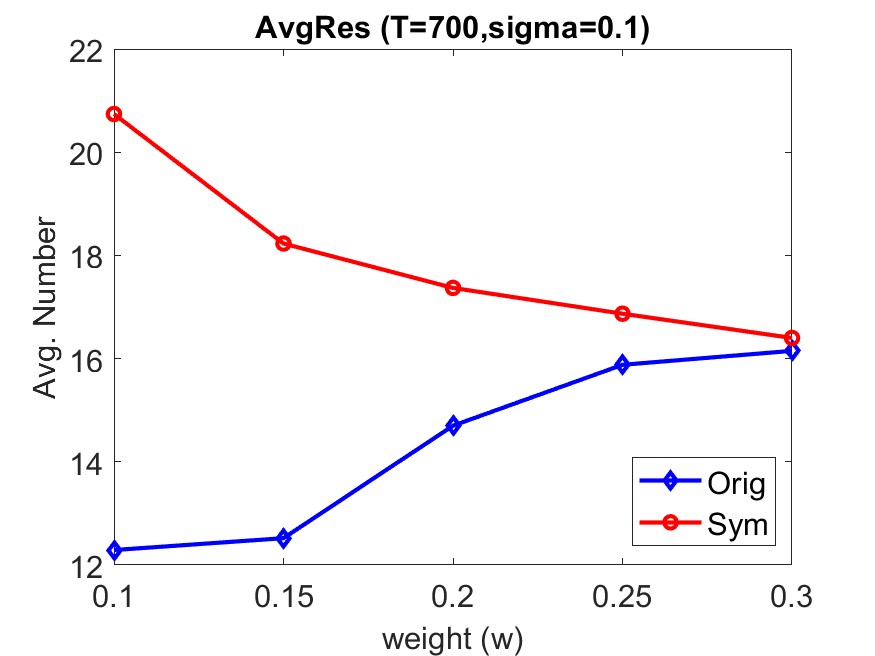}
\includegraphics[width=1.75in]{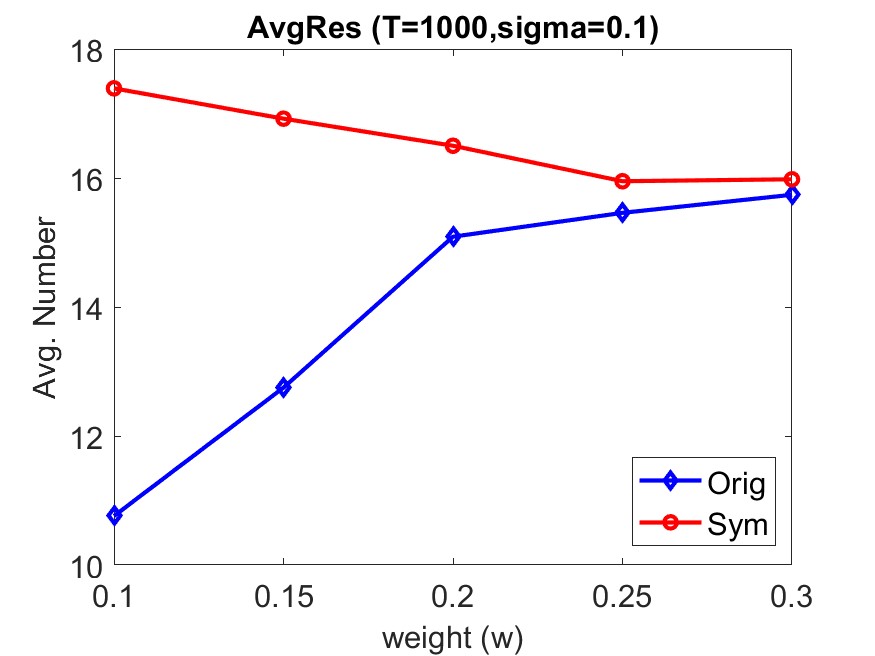}
}
\caption{Average of the estimated number of shared resources $\bar{q}$ ($\sigma = 0.1$).}
\label{fig:avgres-var10}
\end{figure}

In order to examine the issue of false positives further, we also compute the mean of $q$, i.e., the average number of rows in the output $\tilde{\bf A}$, which is the estimated number of shared resources by our algorithm. We denote it by $\bar{q}$ and plot them in Fig.~\ref{fig:avgres-var10}. First, note that except for when $T \leq 500$ and $w_S \leq 0.2$, in most cases $\bar{q}$ does not exceed 20; when the original measurements without symmetrization are used, $\bar{q}$ is mostly below 16, whereas with symmetrized measurements, it is a little higher. For example, when $w_S \geq 0.2$ and $T \geq 700$, since the accuracy is approximately 90 percent or higher, the number of incorrect rows in $\tilde{\bf A}$ is only roughly 3-4 on the average. Together with the earlier observation, this suggests that our algorithm does not produce many incorrect `guesses', and most of the rows in $\tilde{\bf A}$ accurately identify shared resources with the correct sets of sharing NSs when $T \geq 500$ and $w_S \geq 0.2$.

Fig.~\ref{fig:avgres-var10} reveals that, for small $w_S \in$ \{0.1, 0.15\}, $\bar{q}$ is considerably larger when symmetrized measurements are used compared to when original measurements are used. However, this difference is largely due to the fact that the accuracy is much higher with symmetrized measurements. For instance, for $w_S = 0.15$ and $T = 700$, the accuracy for symmetrized measurements is roughly 87 percent, whereas it is approximately 42 percent with original measurements. This translated to roughly 15$\times(0.87 - 0.42) = 6.75$ more correctly identified shared resources in $\tilde{\bf A}$ when symmetrized measurements are used, which is approximately the difference in $\bar{q}$. Therefore, the average number of incorrect rows in $\tilde{\bf A}$ is approximately the same, while the accuracy is much higher with symmetrized measurements. This affirms our intuition regarding symmetrized measurements.

\subsubsection{Robustness of the proposed algorithm}
    \label{subsubsec:Robustness}

In this subsection, we examine the robustness of the proposed algorithm for identifying shared resources against measurement noise. To this end, we vary the noise variance to determine how it affects the accuracy and the number of rows in $\tilde{\bf A}$. 

\begin{figure}[h]
\centerline{
\includegraphics[width=1.75in]{CoveredExactN700-var10.jpg}
\includegraphics[width=1.75in]{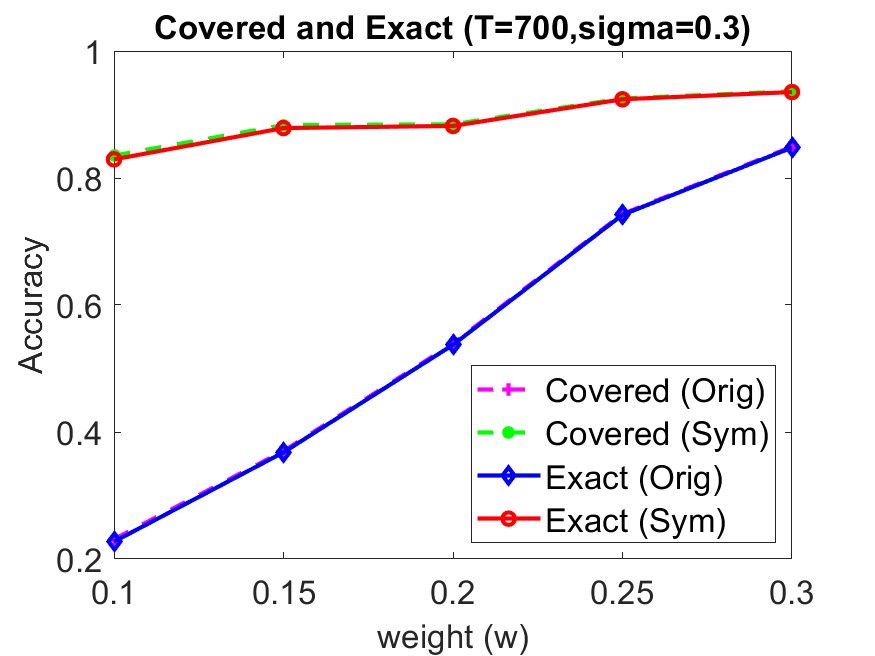}
}
\centerline{
\includegraphics[width=1.75in]{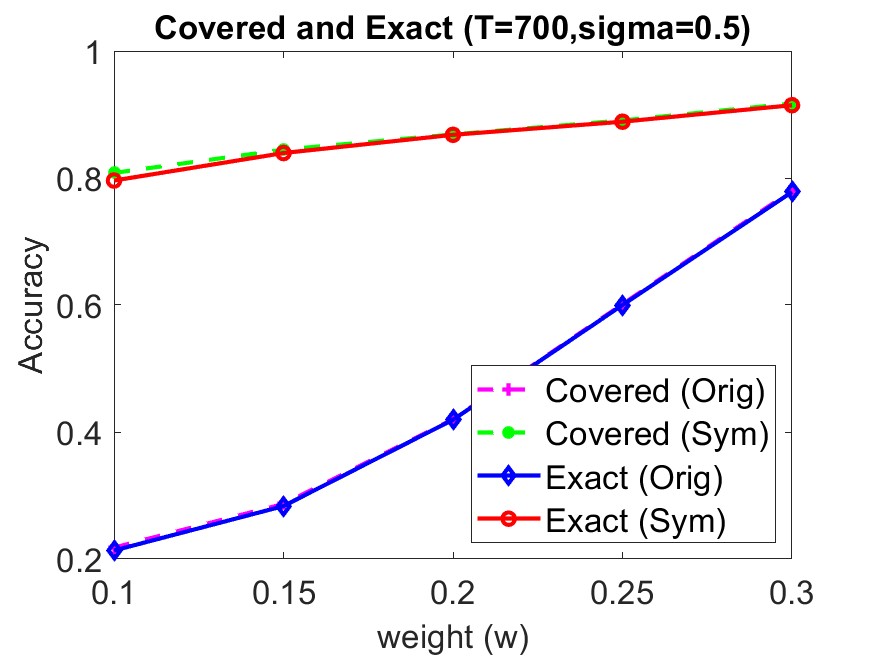}
\includegraphics[width=1.75in]{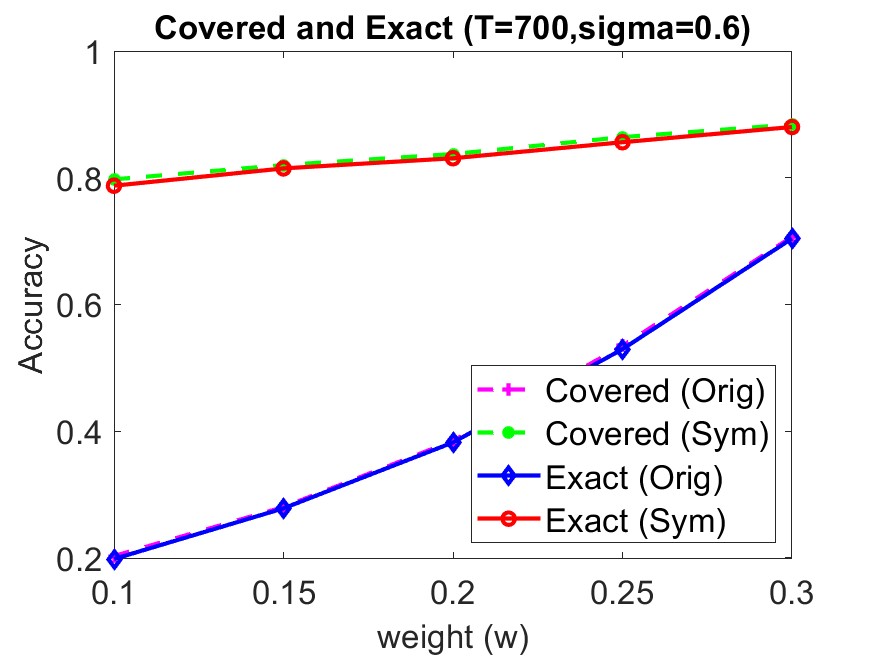}
}
\caption{Accuracy of the proposed algorithm with varying $\sigma$ ($T = 700$).}
\label{fig:exact-varying}
\end{figure}

\begin{figure}[h]
\centerline{
\includegraphics[width=1.75in]{avgResN700-var10.jpg}
\includegraphics[width=1.75in]{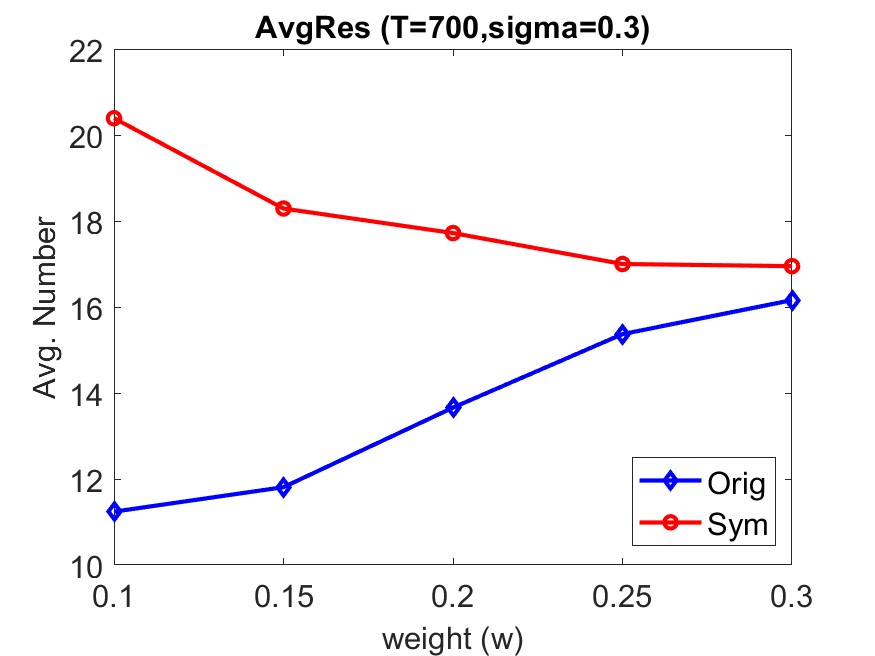}
}
\centerline{
\includegraphics[width=1.75in]{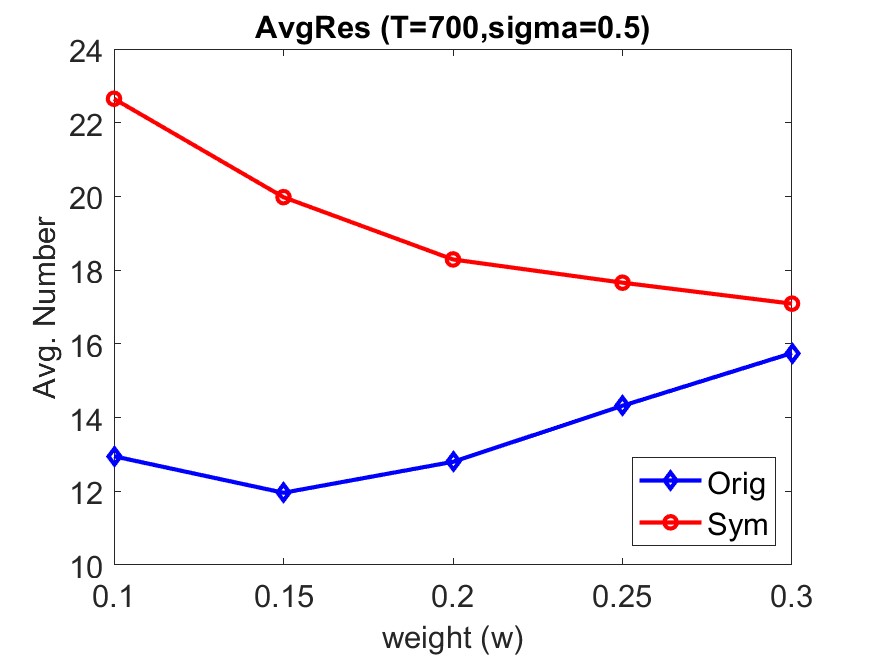}
\includegraphics[width=1.75in]{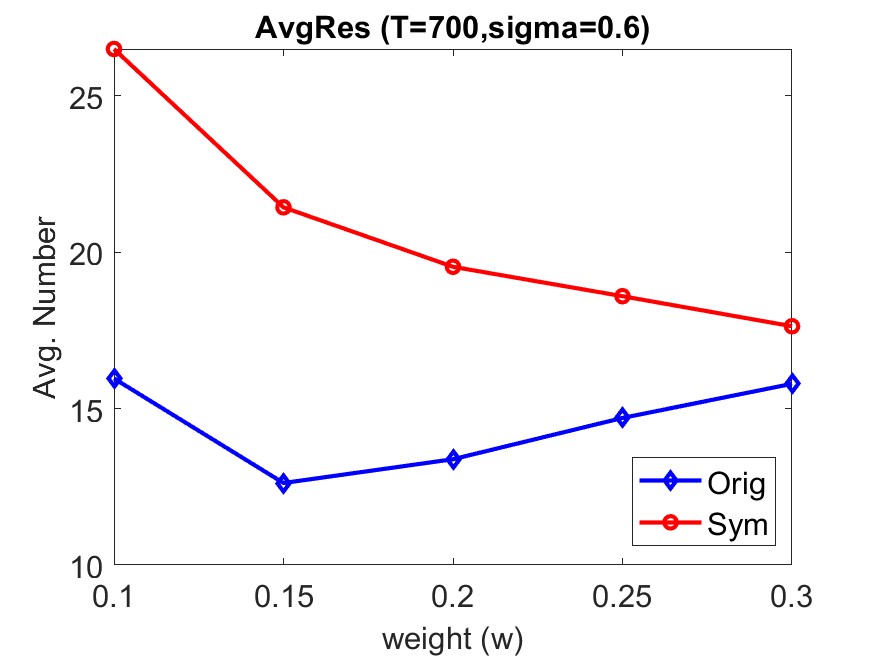}
}
\caption{Average of the estimated number of shared resources, $\bar{q}$,  with varying $\sigma$ ($T = 700$).}
\label{fig:avgres-varying}
\end{figure}

Figs.~\ref{fig:exact-varying} and \ref{fig:avgres-varying} plot the accuracy and $\bar{q}$ (the average number of rows in $\tilde{\bf A}$) as we change the noise variance for $T = 700$. The reported numbers are the average of 100 random runs. The plots show that as the standard deviation $\sigma$ increases from 0.1 to 0.5, the accuracy of the proposed algorithm shown in Fig.~\ref{fig:exact-varying}, especially with symmetrized measurements, changes only little. In addition, the average number of rows in $\tilde{\bf A}$, namely $\bar{q}$, plotted in Fig.~\ref{fig:avgres-varying} increases only slightly over the same interval. On the other hand, when $\sigma$ increases from 0.5 to 0.6, although the decrease in accuracy and the increase in $\bar{q}$ are not large, they are more noticeable especially for smaller $w_S$. The accuracy for $w_S \geq 0.2$ nevertheless remains above 80 percent. Recall that when $\sigma = 0.5$, the standard deviation of the additive noise is 50 percent of the deterministic part of the delay determined by the functions described earlier and, hence, the noise is quite large. Therefore, these findings suggest that our algorithm is robust to measurement noise.

\subsubsection{Detection of misbehaving network slices}
    \label{subsubsec:NumericalMisbehavingNS}

We evaluate the performance of our algorithm for detecting misbehaving NSs that cause greater service interference to other NSs (see Section~\ref{sec:proposed-misbehaving-NSs} for details of the algorithm). For our numerical studies, we select 3 out of 50 NSs (NSs 15, 30, and 45) as misbehaving NSs by changing their utilization levels (associated with the 4 MC states described earlier) to [0.2 0.5 0.7 1.0]; when they are at state 4, their utilization level is 100 percent instead of 90 percent assumed earlier for other NSs. This allows these 3 misbehaving NSs to cause higher congestion at the shared resources. Recall that transition probabilities are selected so that the probability of transitioning to state 4 is usually small. Thus, the 3 misbehaving NSs behave like other normal NSs most of the time. In addition, the difference in utilization levels at MC state 4 is not large and, thus, identifying misbehaving NSs would be challenging. 

\begin{figure}[h]
\centerline{
\includegraphics[width=3.2in, height=1.5in]{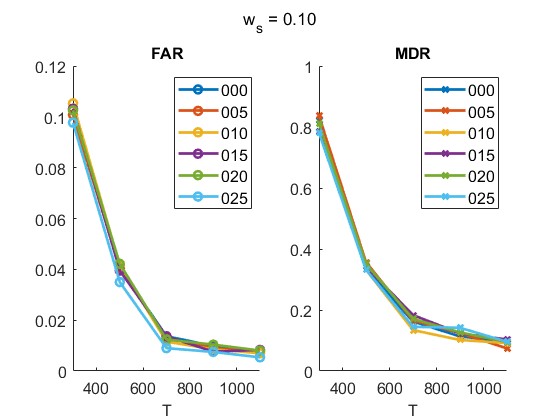}
}
\centerline{
\includegraphics[width=3.2in, height=1.5in]{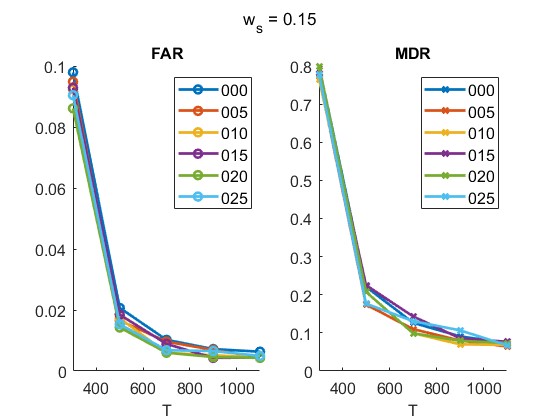}
}
\centerline{
\includegraphics[width=3.2in, height=1.5in]{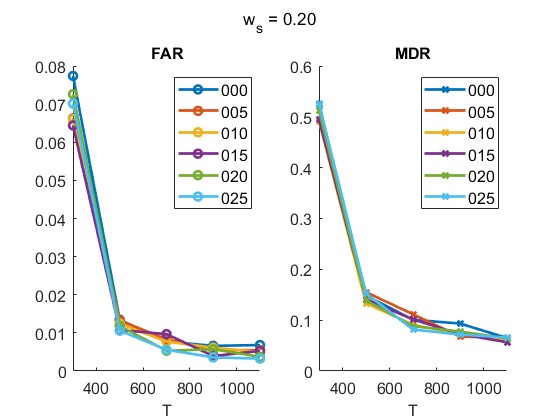}
}
\centerline{
\includegraphics[width=3.2in, height=1.5in]{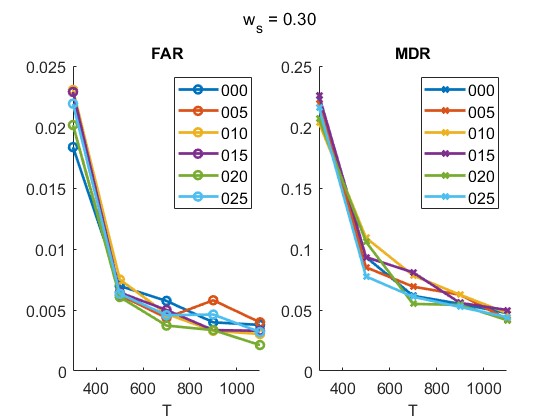}
}
\caption{False alarm rate (FAR) and missed detection rate (MDR) 
    for $w_S \in \{0.10, 0.15, 0.20, 0.30\}$ and $\gamma \in \{0.0, 0.005, 
    0.01, 0.015, 0.02, 0.025\}$.}
\label{fig:misbehaving1}
\end{figure}    

We plot in Fig.~\ref{fig:misbehaving1} the false alarm rate (FAR) and missed detection rate (MDR) as a function of the number of measurements $T$ for different values of weight $w_S$ and threshold $\gamma$. The FAR is the fraction of 47 normal NSs (other than the 3 misbehaving NSs), which are mistakenly identified as misbehaving NSs by our algorithm. The MDR is the fraction of misbehaving NSs, which are not flagged as misbehaving NSs by our algorithm. Here, we only present the numbers with symmetrized measurements, and the reported numbers are the average of 400 random runs. 

First, as expected, except for some fluctuations in the reported numbers, both the FAR and the MDR diminish with the number of measurements $T$. Also, except for when $T = 300$, the FAR is generally small. This suggests that our algorithm rarely flags well-behaved NSs as misbehaving NSs when $T \geq 500$. In addition, the MDR falls quickly as $T$ increases from 300 to 500 and, for $w_S = 0.1$, it drops considerably when $T$ is further increased to 700. It is noteworthy that the MDR for $w_S = 0.2$ and $T \geq 700$ is roughly 10 percent or lower and, for $w_S = 0.3$ and $T = 1100$, the MDR is close to 5 percent. This tells us that even when the interference is weak to moderate (see Fig.~\ref{fig:PCC_w010_020} for the plots of Pearson correlation coefficients), the proposed algorithm can correctly identify most of misbehaving NSs when a sufficient number of measurements are available. 

Second, both the FAR and the MDR drop significantly with increasing $w_S$ when $T$ is small. This is intuitive in that when the number of measurements is limited, the interference should be sufficiently strong to allow the algorithm to correctly identify misbehaving NSs. Thus, when the interference is weak, the algorithm will be unable to differentiate misbehaving NSs from well-behaved NSs and will benefit from stronger interference. This can be easily seen from the reported numbers for $T = 300$; when $w_S \leq 0.2$, the algorithm fails to identify a majority of misbehaving NSs, and only when $w_S = 0.3$, it achieves an MDR of approximately 22 percent. 

Third, recall from Fig.~\ref{fig:SD} that the standard deviations of the measurements for different NSs range from 0.03 to 0.048. For this reason, we select a threshold $\gamma$ between 0 and 0.025.  Somewhat surprisingly the plots indicate that the value of threshold $\gamma$ does not affect the performance of the algorithm significantly as long as the threshold is not too large in relation to the standard deviation of measurements. This seems to suggest that the net benefits of only considering more congested resources with increasing $\gamma$ are offset by the decreasing number of shared resources that are considered.  

Finally, we note that the MDR ranges from 4 to 11 percent when $w_S \geq 0.2$ and $T \geq 700$. Hence, when the interference is weak to at most moderate and sufficiently many measurements are available, the proposed algorithm can correctly identify most of misbehaving NSs with a low FAR ranging only from 0.5 to 1 percent. Therefore, our results in Fig.~\ref{fig:misbehaving1} indicate that the proposed algorithm can be effective at finding misbehaving NSs that can potentially cause harmful interference to other NSs even when the interference is generally weak, provided that a reasonable number of measurements are available.

\subsection{Comparison of Pearson correlation coefficients and Spearman's rank correlation coefficients}
    \label{subsec:CorrCoeff}

As mentioned before, our numerical studies carried out for a wide range settings indicate that Spearman's rank correlation coefficient is better suited for determining pairwise interference among NSs (Stage 1 of the first algorithm described in Section~\ref{sec:Proposed-shared-resources}) than PCC. In order to illustrate this, we plot the distribution of both PCCs and SRCCs for $T=300$ and $T=700$. We select $w_S = 0.3$ and $\sigma = 0.1$ for the example.

\begin{figure}[h]
\centerline{
\includegraphics[width=1.7in]{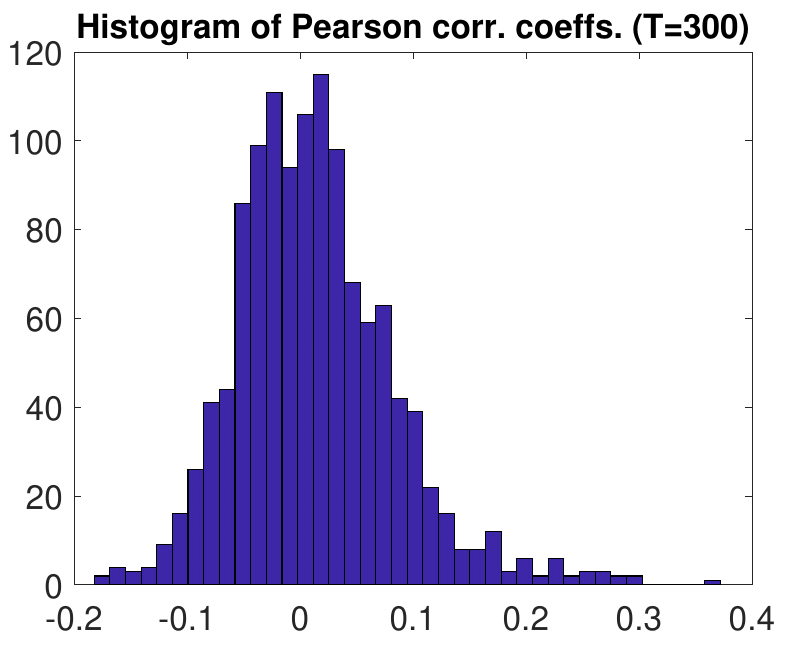}
\includegraphics[width=1.72in]{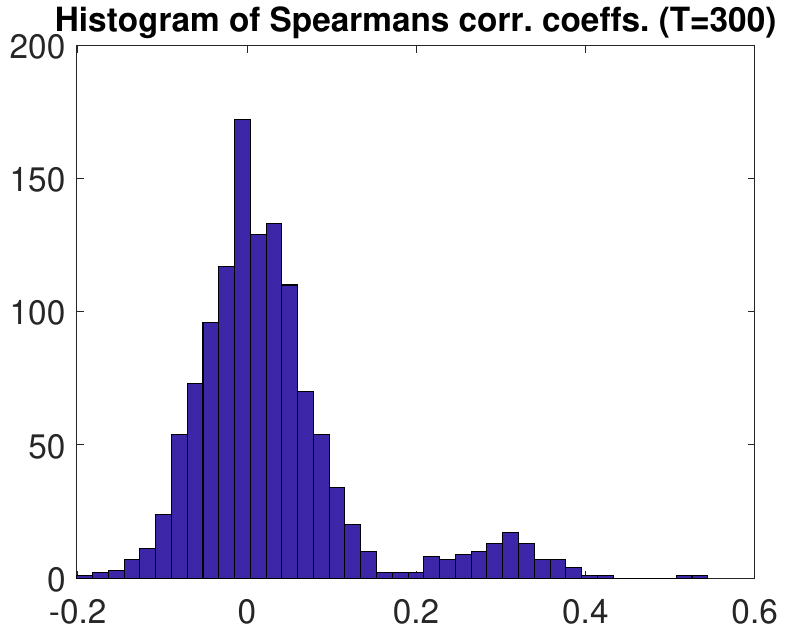}
}
\centerline{(a)}
\centerline{
\includegraphics[width=1.7in]{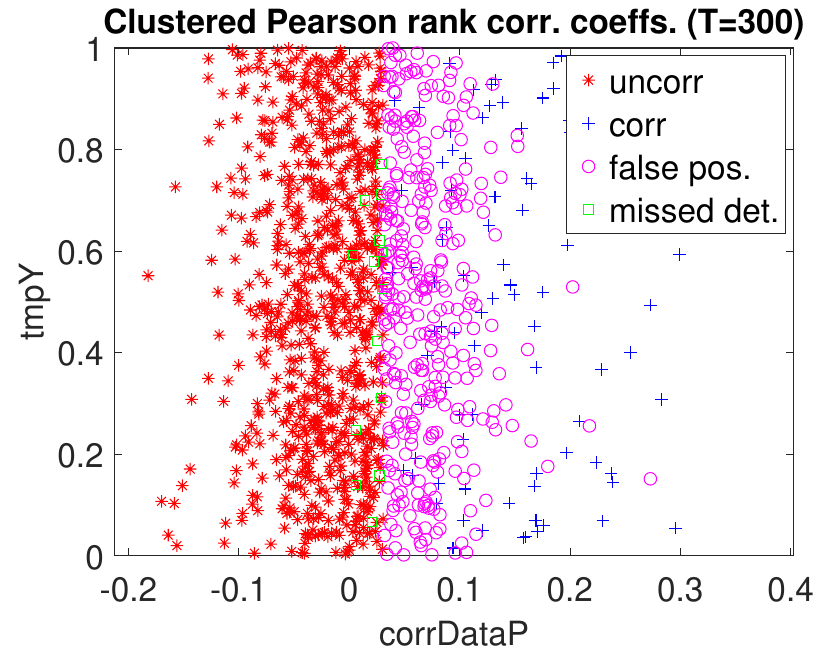}
\includegraphics[width=1.74in]{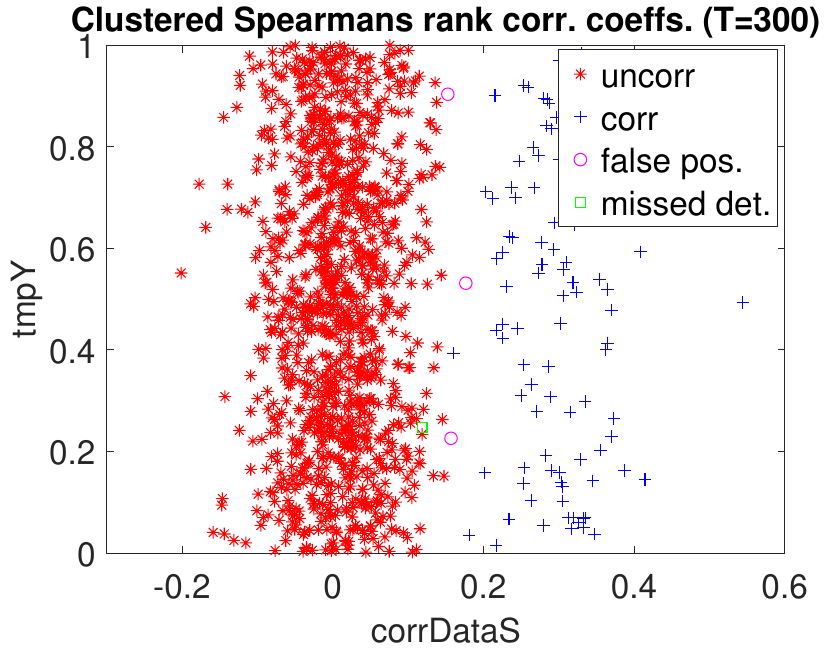}
}
\centerline{(b)}
\caption{Comparison of Pearson correlation coefficients and Spearman's rank correlation coefficient. (a) Histogram of pairwise correlation coefficients, (b) scatter plot of correlation coefficients with label -- \texttt{uncorr}: uncorrelated, \texttt{corr}: correlated, \texttt{false pos.}: false positive, \texttt{missed det.}: missed detection ($T = 300, N = 50, R = 15, \sigma = 0.1, w_S = 0.3$).}
\label{fig:corrCoeff300}
\end{figure}

\begin{figure}[h]
\centerline{
\includegraphics[width=1.7in]{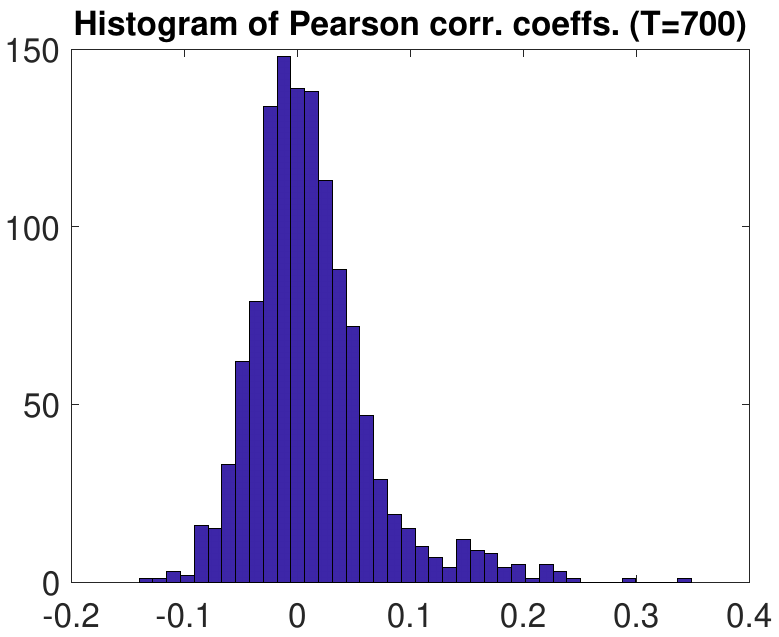}
\includegraphics[width=1.72in]{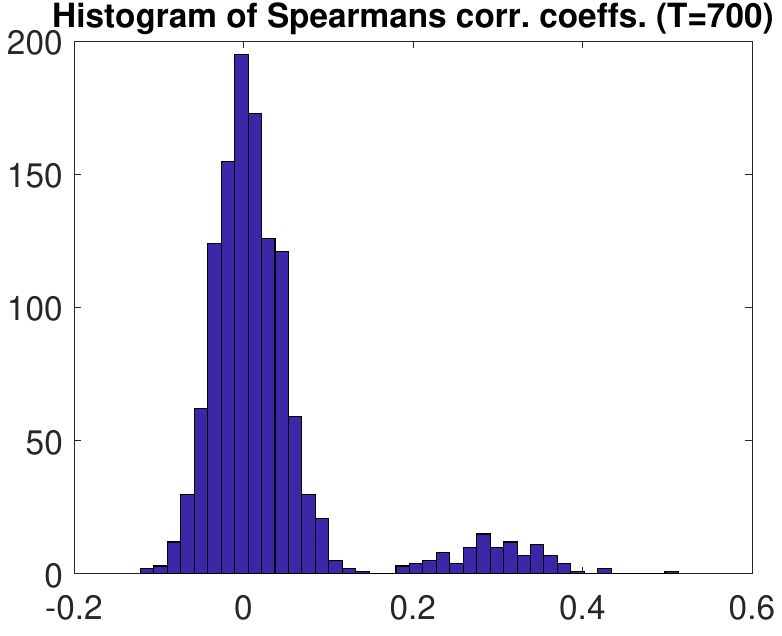}
}
\centerline{(a)}
\centerline{
\includegraphics[width=1.7in]{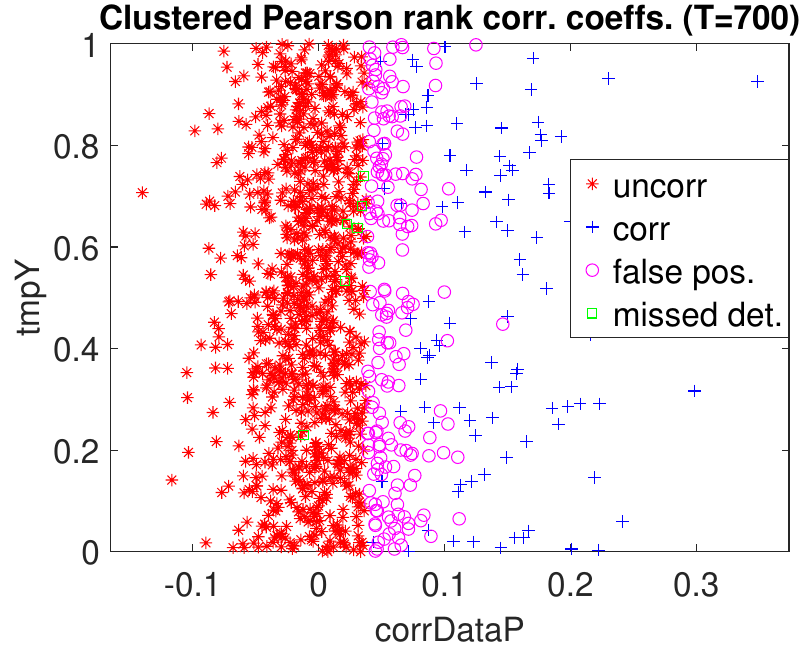}
\includegraphics[width=1.77in]{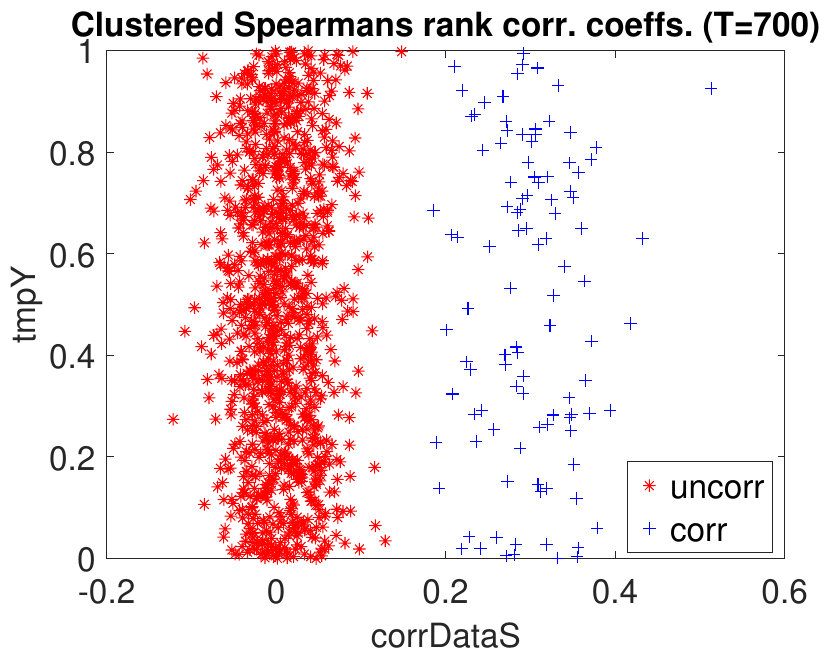}
}
\centerline{(b)}
\caption{Comparison of Pearson correlation coefficients and Spearman's rank correlation coefficient. (a) Histogram of pairwise correlation coefficients, (b) scatter plot of correlation coefficients with label -- \texttt{uncorr}: uncorrelated, \texttt{corr}: correlated, \texttt{false pos.}: false positive, \texttt{missed det.}: missed detection ($T = 700, N = 50, R = 15, \sigma = 0.1, w_S = 0.3$).}
\label{fig:corrCoeff700}
\end{figure}

Recall that in Stage 1 of the algorithm for identifying shared resources, we are interested in identifying all pairwise interference between two NSs sharing at least one resource. A {\em false positive} happens when the interference graph includes an edge between two NSs that do not share any resource. A {\em missed detection} refers to a missing edge in the interference graph between two NSs that share a resource. 

Figs.~\ref{fig:corrCoeff300} and \ref{fig:corrCoeff700} plot the histograms of PCCs and SRCCs and their scatter plots for $T=300$ and $T=700$, respectively, where the $y$-coordinate values (\texttt{tmpY}) are i.i.d. Uniform(0,1) random variables. First, it is obvious from the plots that the SRCCs between NSs sharing a resource (\texttt{corr} and \texttt{missed det.} in the plots) are better separated from those of the NSs that do not share any resource (\texttt{uncorr} and \texttt{false pos.}) than with the PCCs; when two NSs share a resource, their SRCC tends to be larger (with the mean just below 0.3) than their PCC (with the mean around 0.15). 

Second, the histograms of SRCCs suggest a mixture distribution -- one (roughly Gaussian) distribution centered around zero models the SRCCs between NSs that do not share a resource, and the other distribution with a smaller weight is centered around 0.3. However, the histogram of PCCs for $T = 300$ appears to have only single component distribution without a discernible second component, i.e., not a mixture distribution, while for $T = 700$, there appear to be two components - the second component centered just below 0.2. This becomes more evident when we look at the scatter plots. A closer examination reveals that there are many false positives and missed detections in the scatter plots of PCCs even for $T = 700$, while that of SRCCs has no missed detection or false positive for $T = 700$. This clearly illustrates that SRCC is a better choice for detecting pairwise interference among NSs.


\section{Conclusion and Future Direction}    \label{sec:Conclusion}

We investigated the problem of identifying potential service interference among NSs that share physical or virtual resources. The first part of the problem is formulated as one of finding a set of cliques, where each clique represents a group of NSs that share a resource. We proposed a new algorithm based on factor analysis. It makes use of pairwise correlations among NSs to construct an interference graph, which is then used to identify a subset of cliques. Furthermore, the output of the algorithm tells us how the state of each shared resource affects the performance of NSs that share the resource. 

The second part of the problem requires estimating the congestion levels at shared resources and relative congestion of each NS compared to those of shared resources. Based on a key observation that the specific factors behave differently between normal NSs and misbehaving NSs, we proposed a novel algorithm that can identify misbehaving NSs with weak to moderate interference among NSs. Numerical results show that, when sufficient measurements are available, our algorithm can correctly identify most of the shared resources in the networks along with the subset of NSs that share each identified resource and misbehaving NSs that can cause potentially adverse interference to other well-behaved NSs. 


\section*{Acknowledgment \& Disclaimer}
We thank Junxiao Shi and Davide Pesavento at NIST for their help with the testbed. 

Certain equipment, instruments, software, or materials, commercial or non-commercial, are identified in this paper in order to specify the experimental setup adequately. Such identifications do not imply a recommendation or endorsement of any product or service by NIST, and do not imply that they are the best available for the purpose.

\bibliographystyle{plain}
\bibliography{ref}

@misc{open5gcore,
  title        = {Open5GCore},
  author       = {Fraunhofer FOKUS},
  year         = {2025},
  howpublished = {\url{https://www.open5gcore.org}}
}

@MISC{iperf,
title        = {Iperf2: A means to measure network responsiveness and throughput},
howpublished = {\url{https://sourceforge.net/projects/iperf2/}},
year         = {2025}
}

@article{ditg,
author    = {Alessio Botta and	Alberto Dainotti and Antonio Pescap{\`e}},
title     = {A tool for the generation of realistic network workload for emerging networking scenarios},
journal   = {Computer Networks},
volume    = {56},
number    = {15},
year      = {2012},
pages     = {3531-3547},
howpublished = {\url{https://www.traffic.comics.unina.it/software/ITG/}}
}

@article{zhao2008ml,
  title={{ML} estimation for factor analysis: {EM} or non-{EM}?},
  author={Zhao, J-H and Yu, Philip LH and Jiang, Qibao},
  journal={Statistics and computing},
  volume={18},
  pages={109--123},
  year={2008},
  publisher={Springer}
}

@article{dempster1977maximum,
  title={Maximum likelihood from incomplete data via the {EM} algorithm},
  author={Dempster, Arthur P and Laird, Nan M and Rubin, Donald B},
  journal={Journal of the royal statistical society: series B (methodological)},
  volume={39},
  number={1},
  pages={1--22},
  year={1977},
  publisher={Wiley Online Library}
}

@article{eppstein2013listing,
author = {Eppstein, David and L\"{o}ffler, Maarten and Strash, Darren},
title = {Listing All Maximal Cliques in Large Sparse Real-World Graphs},
year = {2013},
issue_date = {2013},
publisher = {Association for Computing Machinery},
address = {New York, NY, USA},
volume = {18},
issn = {1084-6654},
url = {https://doi.org/10.1145/2543629},
doi = {10.1145/2543629},
journal = {ACM J. Exp. Algorithmics},
month = nov,
articleno = {3.1},
numpages = {21}
}

@book{BarthKnott, 
    title={Latent Variable Models and Factor Analysis: Kendall's Library of Statistics 7},
    edition={2nd},
    author={David J. Bartholomew and Martin Knott}, 
    year={1999},
    publisher={Wiley}
}

@book{Daniel,
    title={Applied Nonparametric Statistics}, 
    edition={2nd},
    author={Wayne W. Daniel}, 
    year={1989},
    publisher={Wadsworth Publishing Company}
}

@inproceedings{jain2017systematic,
  title={A systematic review of nature inspired load balancing algorithm in heterogeneous cloud computing environment},
  author={Jain, Pankaj and Sharma, Sanjay Kumar},
  booktitle={2017 conference on information and communication technology (CICT)},
  pages={1--7},
  year={2017}
}

@inproceedings{zhang2019adaptive,
  title={Adaptive interference-aware VNF placement for service-customized 5G network slices},
  author={Zhang, Qixia and Liu, Fangming and Zeng, Chaobing},
  booktitle={IEEE INFOCOM 2019-IEEE Conference on Computer Communications},
  pages={2449--2457},
  year={2019}
}

@article{zambianco2020interference,
  title={Interference minimization in 5G physical-layer network slicing},
  author={Zambianco, Marco and Verticale, Giacomo},
  journal={IEEE Transactions on Communications},
  volume={68},
  number={7},
  pages={4554--4564},
  year={2020}
}

@inproceedings{xu2013rethinking,
  title={Rethinking virtual machine interference in the era of cloud applications},
  author={Xu, Tianni and Sui, Xiufeng and Yao, Zhicheng and Ma, Jiuyue and Bao, Yungang and Zhang, Lixin},
  booktitle={2013 IEEE 10th International Conference on High Performance Computing and Communications \& 2013 IEEE International Conference on Embedded and Ubiquitous Computing},
  pages={190--197},
  year={2013}
}

@inproceedings{amri2017inter,
  title={Inter-VM interference in cloud environments: A survey},
  author={Amri, Sabrine and Hamdi, Hedi and Brahmi, Zaki},
  booktitle={2017 IEEE/ACS 14th International Conference on Computer Systems and Applications (AICCSA)},
  pages={154--159},
  year={2017}
}

@inproceedings{coates2000,
  title={Network loss inference using unicast end-to-end measurement},
  author={Coates, Mark and Nowak, Robert},
  booktitle={ITC Seminar on IP Traffic, Measurement and Modelling},
  pages={915-923},
  year={2001}
}

@inproceedings{duffield2001,
  title={Inferring link loss using striped unicast probes},
  author={Duffield, Nick G. and Lo Presit, Francesco and Paxson, Vern and Towsley, Don},
  booktitle={IEEE INFOCOM},
  pages={915-923},
  year={2001}
}

@inproceedings{malekzadeh2013,
  title={Network topology inference from end-to-end unicast measurements},
  author={Malekzadeh, Amir and MacGregor, Mark H.},
  booktitle={27th International Conference on Advanced Information Networking and Applications Workshops},
  pages={1101-1106},
  year={2013}
}

@inproceedings{kambadur2012measuring,
  title={Measuring interference between live datacenter applications},
  author={Kambadur, Melanie and Moseley, Tipp and Hank, Rick and Kim, Martha A},
  booktitle={SC'12: Proceedings of the International Conference on High Performance Computing, Networking, Storage and Analysis},
  pages={1--12},
  year={2012}
}

@inproceedings{ziot2001,
  title={Estimation of network link loss rates via chaining in multicast trees},
  author={Ziotopoulos, Ayis and Hero, Alfred and Wasserman, K.M.},
  booktitle={IEEE International Conference on Acoustics, Speech, and Signal Processing},
  pages={2517-2520},
  year={2001}
}

@inproceedings{bestavros2002,
  title={Inference and labeling of metric-induced network topologies},
  author={Bestavros, Azer and Byers, John W. and Harfoush, Khaled},
  booktitle={IEEE INFOCOM},
  pages={628-637},
  year={2002}
}

@inproceedings{coates2002a,
  title={Maximum likelihood network topology identification from edge-based unicast measurements},
  author={Coates, Mark and Castro, Rui and Nowak, Robert and Gadhiok, Manik and King, Ryan and Tsang, Yolanda},
  booktitle={ACM Sigmetrics},
  pages={11-20},
  year={2002}
}

@inproceedings{duffield2001a,
  title={Multicast topology inference from measured end-to-end loss},
  author={Duffield, Nick G. andd Horowitz, Joseph and Lo Presti, Francesco and Towsley, Don},
  booktitle={IEEE INFOCOM},
  pages={1636-1645},
  year={2001}
}

@inproceedings{ciss2025,
  title={{Detection of performance interference among network slices in 5G/6G systems}},
  author={Mai, Van-Sy and La, Richard J. and Zhang, Tao},
  booktitle={Annual Conference on Information Science and Systems},
  pages={},
  year={2025}
}

@article{Foukas2017,
  title={{Network Slicing in 5G\: Survey and Challenges}},
  author={Foukas, Xenofon and Patounas, Georgios and Elmokashfi, Ahmed and Marina, Mahesh M.},
  journal={IEEE Communications Magazine},
  pages={94-100},
  volume={55}, 
number={5}, 
  month={May}, 
  year={2017}
}

@article{caceres1999,
  title={{Multicast-based inference of network-internal loss characteristics}},
  author={Caceres, Ramon and Duffield, Nick G. and Horowitz, Joseph and Towsley, Don},
  journal={IEEE Transactions on Information Theory},
  pages={2462-2480},
  volume={45}, 
number={y}, 
  month={November}, 
  year={1999}
}

@article{Tsang2003,
  title={{Network delay tomography}},
  author={Tsang, Yolanda and Coates, Mark J. and Nowak, Roboert},
  journal={IEEE Transactions on Signal Processing},
  pages={2125-2136},
  volume={51}, 
    number={8}, 
  month={August}, 
  year={2003}
}

@article{duffield2002a,
  title={{Multicast topology inference from measured end-to-end loss}},
  author={Duffield, Nick G. and Horowitz, Joseph and Lo Presti, Francesco and Towsley, Don},
  journal={IEEE Transactions on Information Theory},
  pages={26-45},
  volume={48}, 
    number={2}, 
  month={January}, 
  year={2002}
}

@article{Rost2017,
  title={{Nework Slicing to Enable Scalability and Flexibility in 5G
            Mobile Networks}},
  author={Rost, Peter and Mannweiler, Christian and Michalopoulos,      
          Diomidis S. and Sartori, Cinzia and Sciancalepore, Vincenzo 
          and Sastry, Nishanth},
  journal={IEEE Communications Magazine},
  pages={72-79},
  volume={55}, 
number={5}, 
  month={May}, 
  year={2017}
}

@article{ArXiV,
  title={{Detection of performance interference among network slices in 
         5G/6G systems}},
  author={Mai, Van Sy and La, Richard J. and Zhang, Tao},
  journal={arXiv:2412.01584},
  year={2024}
}

\end{document}